\documentclass[pra,aps,10pt,longbibliography,twocolumn,superscriptaddress]{revtex4-2}% 
\usepackage{amsmath}
\usepackage{amssymb}
\usepackage{amsthm}
\usepackage{amsfonts}
\usepackage{enumerate}
\usepackage{latexsym}
\usepackage{bm}
\usepackage{graphicx}
\usepackage{subfigure}
\usepackage{color}
\usepackage[dvipsnames]{xcolor}
\usepackage[colorlinks]{hyperref} % http://ctan.org/pkg/hyperref
\hypersetup{
 colorlinks,
 citecolor=Violet,
 linkcolor=Red,
 urlcolor=Blue}
\usepackage{ulem}%using \sout from the ulem package
\usepackage{cancel}
\usepackage{float}

\newcommand{\beq}{\begin{equation}}
\newcommand{\eneq}{\end{equation}}
\input{epsf}

\definecolor{awesome}{rgb}{1.0, 0.13, 0.32}
\definecolor{electricpurple}{rgb}{0.75, 0.0, 1.0}

\begin{document}

\title{
Enhancement of vibrationally assisted energy transfer by proximity to exceptional points, probed by fluorescence-detected vibrational spectroscopy
}

\author{Zeng-Zhao Li}
\email{zengzhaoli09@gmail.com}
\affiliation{Department of Chemistry, University of California, Berkeley, California 94720, USA}
\affiliation{Berkeley Center for Quantum Information and Computation, Berkeley, California 94720, USA}
\affiliation{Center for Quantum Coherent Science, University of California, Berkeley, California 94720, USA}

\author{K. Birgitta Whaley}
\email{whaley@berkeley.edu}
\affiliation{Department of Chemistry, University of California, Berkeley, California 94720, USA}
\affiliation{Berkeley Center for Quantum Information and Computation, Berkeley, California 94720, USA}
\affiliation{Center for Quantum Coherent Science, University of California, Berkeley, California 94720, USA}

\begin{abstract}
Emulation of energy transfer processes in natural systems on quantum platforms can further our understanding of complex dynamics in nature. One notable example is the demonstration of vibrationally assisted energy transfer (VAET) on a trapped-ion quantum emulator, which offers insights for the energetics of light harvesting. In this work, we expand the study of VAET simulation with trapped ions to a non-Hermitian quantum system comprising a ${\cal PT}$-symmetric chromophore dimer weakly coupled to a vibrational mode. We first characterize exceptional points (EPs) and non-Hermitian features of the excitation energy transfer processes in the absence of the vibration, finding a degenerate pair of second-order EPs. 
Exploring the non-Hermitian dynamics of the whole system including vibrations, we find that energy transfer accompanied by absorption of phonons from a vibrational mode can be significantly enhanced near such a degenerate EP. Our calculations reveal a unique spectral feature accompanying the coalescing of eigenstates and eigenenergies that provides a novel approach to probe the degenerate EP by ﬂuorescence-detected vibrational spectroscopy. Enhancement of the VAET process near the EP is found to be due to maximal favorability of phonon absorption at the degenerate EP, enabling multiple simultaneous excitations.  
Our work on improving VAET processes in non-Hermitian quantum systems paves the way for leveraging non-Hermiticity in quantum dynamics related to excitation energy transfer.
\end{abstract}
\date{\today}
\maketitle
%\tableofcontents

\section{Introduction}

Non-Hermitian Hamiltonians respecting parity-time (${\cal PT}$) symmetry have been intensively studied over the past two decades~\cite{Bender19,El-GanainyChristodoulides18nphys,Ozdemir19nmat,ZhaoFeng18nsr,AshidaGongUeda20advphys}. A characteristic feature of non-Hermitian ${\cal PT}$-symmetric systems is a spectral degeneracy known as the exceptional point (EP) at which 
there is both a degeneracy in eigenenergies and coalescence of eigenvectors. 
The non-Hermitian physics due to this unique feature has been explored in many classical systems~\cite{Dembowski01prl,RuterKip10nphys,HodaeiKhajavikhan14science}, with a wide range of applications such as enhanced sensing ~\cite{Wiersig14prl,ChenYang17nature,HodaeiKhajavikhan17nature}, laser emission management~\cite{BrandstetterRotter14ncomms,PengOzdemirYang16pnas,WongZhang16nphoton}, and wave transport control~\cite{XuHarris16nature,DopplerRotter16nature,ZhangChan18prx}. 
As interest grows also in exploring non-Hermitian physics in the quantum realm, 
various approaches such as Hamiltonian dilation~\cite{WuRongDu19science} and dissipation engineering~\cite{NaghilooMurch19nphys} have been exploited to investigate non-Hermitian quantum physics. EPs have been observed in simple systems realized experimentally on platforms such as NV centers~\cite{WuRongDu19science}, superconducting circuits~\cite{NaghilooMurch19nphys,Rotter19nphys}, trapped ions~\cite{WangJingChen21pra,DingZhang21prl,quinn2023observing}, and ultracold atoms~\cite{LiJoglekarLuo19ncommun} and extension to larger systems appears feasible with NMR systems~\cite{GautamDoraiArvind22arXiv}. 
Recent developments include an investigation of the effects of quantum jumps on non-Hermitian dynamics~\cite{ChenMurch21arXiv}, emergence of ${\cal PT}$-symmetry in open quantum systems~\cite{HuberRabl20}, exceptional points in Floquet systems~\cite{LiJoglekarLuo19ncommun,AbbasiMurch21arXiv}, optimal control of non-Hermitian qubits~\cite{LewalleWhaley22}, and entanglement speedup in proximity to high-order exceptional points~\cite{LiChenWhaley22prl}. 
In addition, investigation of non-Hermitian physics of topological systems~\cite{ZhuChen14pra,Lieu18prb,LiWhaley21arXiv} has shown that topological phases can be enriched by non-Hermiticity~\cite{GongAshidaUeda18prx,YaoWang18prl,KawabataUedaSato19prx,Bergholtz21rmp}.

While considerable effort has been dedicated to exploring nontrivial properties of non-Hermitian quantum systems, little attention has been given to applications in the area of excitation energy transfer. One important issue in this field is understanding the observed long coherence in photosynthetic light-harvesting systems, which is generally agreed to rely on an interplay between excitonic and vibrational degrees of freedom~\cite{CinaFleming04jpca,Christensson12jpcb,PlenioHuelga13jcp}. 
With the development of controllable quantum platforms for simulating quantum phenomena in nature~\cite{PotocnikWallraff18ncomm,GormanHaeffner18prx}, it has become possible to engineer vibrationally assisted energy transfer (VAET) in a trapped-ion quantum simulator, allowing analysis of uphill energy transfer processes~\cite{GormanHaeffner18prx}. This enabled characterization of the interplay between vibration-assisted and environment-assisted energy transfer~\cite{LiSarovarWhaley22njp} as well as identification of collective behaviors of vibrations that can give rise to novel mechanisms such as heteroexcitation~\cite{LiSarovarWhaley21njp}. 
However, all demonstrations of VAET so far have focused on Hermitian systems.
This raises the question of whether the VAET processes are achievable in non-Hermitian quantum systems. If so, it is important to identify what new aspects non-Hermiticity can bring to excitation energy transfer, as well as to determine whether the non-Hermitian VAET might offer advantages over its Hermitian counterpart.

To shed light on these questions and explore new research directions in non-Hermitian excitation energy transfer, we study here effects of the non-Hermiticity on the transfer processes assisted by vibrations in a ${\cal PT}$-symmetric chromophore dimer, representing a simple VAET system relevant to light harvesting dynamics of photosynthetic systems. 
We first characterize the EPs of the chromophore dimer in the absence of vibrations, identifying the non-Hermitian features of oscillations induced by gain and loss and non-equilibrium steady states of excitation energy transfer in the ${\cal PT}$-symmetric unbroken and broken phase, respectively. 
The crucial characteristic of these EPs lies in their classification as second-order EPs that are two-fold degenerate, i.e., each parameter set yielding two distinct second-order EPs.  
This has significant implications for both the non-Hermitian dynamics and the dimer spectroscopy. 
By then considering the non-Hermitian dynamics of the entire system, including both the ${\cal PT}$-symmetric dimer and a weakly coupled vibration, we show that the energy transfer processes involving absorption of phonons from the vibration can be significantly enhanced near such a degenerate EP of the ${\cal PT}$-symmetric dimer. This enhancement stands in sharp contrast to the relatively low transfer efficiency observed in the corresponding Hermitian case. 

In addition to this significant enhancement of energy transfer near a degenerate EP, we also observe a unique feature in fluorescence-detected vibrational spectroscopy of the dimer that accompanies the coalescence of eigenstates or eigenenergies of the non-Hermitian dimer at the EP. 
Observation of this feature constitutes a witness of the EP and ${\cal PT}$-symmetry phase transition, suggesting a novel spectroscopic approach that can further be used to probe the distance of the system from an EP. 

We show here that the significant enhancement of VAET can be attributed to the increasingly favorable absorption of phonons as the degenerate EP is approached. This EP degeneracy enables multiple simultaneous excitations between  
eigenstates associated with distinct EPs 
to be supported. 
We analyze the robustness of our results by examining the response of the enhanced VAET processes to varying strength of vibrational coupling, temperature, and decoherence effects.

The remainder of the paper is structured as follows. In Sec.~\ref{sec:model}, we present the model of a ${\cal PT}$-symmetric chromophore dimer weakly coupled to a vibrational mode. In Sec.~\ref{sec:dimer}, we demonstrate the existence of EPs in the dimeric system without coupling to the vibration and analyze the non-Hermitian features observed in the excitation energy transfer processes. In Sec.~\ref{sec:VAET_spectrum}, we investigate the dynamical and spectral features of the enhanced VAET processes and analyze the phonon-absorption mechanism. The robustness of the results is discussed in Sec.~\ref{sec:robustness}. Finally, in Sec.~\ref{sec:conclusion} we summarize our results and discuss experimental feasibility of realizing this model with a trapped ion simulator.

\begin{figure}[hbt!]
\centering
  \includegraphics[width=.99\columnwidth]{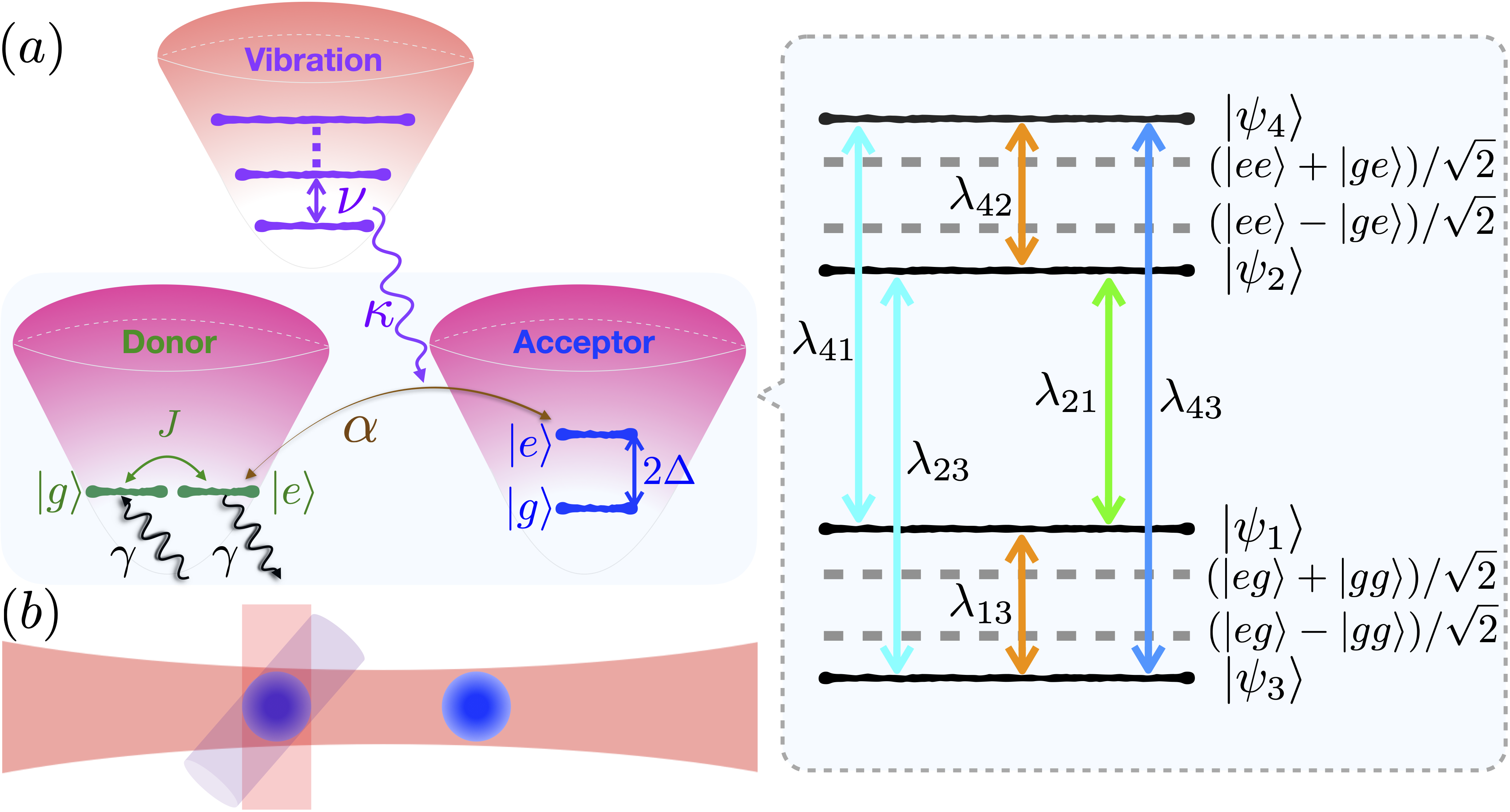}
\caption{(color online) (a) Left panel: Schematic of a non-Hermitian chromophore donor-acceptor dimer coupled to a vibration. The dimer consists of a donor (in green) with a tunneling coupling strength $J$ and gain-loss rate $\gamma$ for its two equal energy states, and an acceptor (in blue) with an energy difference of $2\Delta$. The vibration (in purple) with a frequency $\nu$ is coupled to the donor with a coupling strength $\kappa$. Right panel: Eigenstates of the non-Hermitian dimer and its Hermitian limit. Solid black lines represent the eigenstates $|\psi_i\rangle$
of the non-Hermitian dimer [eigenstates of Eq.~(\ref{eq:H}) with $\kappa=0$] and grey dashed lines represent eigenstates of the corresponding uncoupled Hermitian dimer [eigenstates of Eq.~(\ref{eq:H_dimer}) with $\gamma=\alpha=0$]. 
For the non-Hermitian dimer, at the EP where the frequency differences $\lambda_{13(42)}$ approach zero, a degeneracy occurs in each of two subspaces spanned by states $|\psi_{1}\rangle$ and $|\psi_{3}\rangle$, and  $|\psi_{2}\rangle$ and $|\psi_{4}\rangle$, respectively, which is in both cases accompanied by coalescence of the corresponding eigenvector pair, i.e.,   $|\psi_{1}\rangle$ and $|\psi_{3}\rangle$ in the former, as well as $|\psi_{2}\rangle$ and $|\psi_{4}\rangle$ in the latter.
This EP is thus a two-fold degenerate second-order EP. Also at this point, all other transition frequencies become identical. 
Up-down arrows between pairs of eigenstates $\{|\psi_i\rangle,\,|\psi_j\rangle\}$ indicate associated transitions with frequency difference $\lambda_{ij}(\equiv\lambda_i-\lambda_j)$. 
These transitions are enhanced by resonant coupling to the vibrational mode $\nu$, enabling VAET \cite{GormanHaeffner18prx}.
The schematic diagram presented in the right panel is constructed for the parameter values $\Delta/J=8$, $\alpha/J=1$, and $\kappa/J=0$. 
Schematic of the ion trap and laser beams envisaged in an experimental realization. Two ions are confined within the trap. A global laser beam along the axis of the trap facilitates the interaction responsible for creating the donor-acceptor coupling. Another tightly focused laser beam is localized on the donor ion, enabling single-ion quantum state manipulation and generating the site-vibration coupling. Additional laser drives are utilized for laser heating and cooling, 
to manipulate the gain and loss parameters on the donor ion.
}
\label{fig:schematic}
\end{figure}

\section{VAET dimer model \label{sec:model}}

A typical system for demonstrating the VAET phenomenon is a chromophore dimer weakly coupled to a vibrational mode, illustrated in the left panel of Fig.~\ref{fig:schematic}(a). Adding gain and loss, the system can be  described by the Hamiltonian (setting $\hbar=1$)
\begin{eqnarray}
H &=& 
 J\sigma_x^{(d)}  -i\gamma \sigma_z^{(d)}
  + \alpha\sigma_x^{(d)}\sigma_x^{(a)} 
  + \Delta\sigma_z^{(a)} \notag\\
&& 
+  \kappa\sigma_z^{(d)}(a+a^{\dagger}) 
 + \nu a^{\dagger}a .
\label{eq:H}
\end{eqnarray}
Here the Pauli operators $\sigma_r^{(j)}$ ($r=x, y, z$) are defined with respect to the two lowest levels ($|g\rangle_j$ and $|e\rangle_j$) of the donor ($j=d$) or acceptor ($j=a$) chromophore site, i.e., 
% $\sigma_x^{(d)}=|g\rangle_d\langle e| + |e\rangle_d\langle g|$, $\sigma_z^{(d)}=|e\rangle_d\langle e| - |g\rangle_d\langle g|$, $\sigma_z^{(a)}=|e\rangle_a\langle e| - |g\rangle_a\langle g|$, and $\sigma_x^{(a)}=|e\rangle_a\langle g| + |g\rangle_a\langle e|$
$\sigma_x^{(j)}=|g\rangle_j\langle e| + |e\rangle_j\langle g|$ and $\sigma_z^{(j)}=|e\rangle_j\langle e| - |g\rangle_j\langle g|$ where $j=d,a$. 
The parameters in the Hamiltonian are as follows: the two equal energy levels of the donor have gain-loss rates $\gamma$ and are coupled with a tunneling of strength $J$, the acceptor has an energy gap $2\Delta$, $\alpha$ represents the excitonic coupling between the donor and acceptor, and $\kappa$ denotes the coupling strength of the donor to a vibrational mode with frequency $\nu$. 
This Hamiltonian is specifically tailored for ion trap emulations of energy transfer, which have proven to be a successful platform for demonstrating the VAET phenomenon~\cite{GormanHaeffner18prx}. We note that this Hamiltonian design differs in some respects from that of a natural chromophore system~\cite{LiSarovarWhaley22njp}.

The VAET phenomenon has already been experimentally engineered on a Hermitian trapped-ion platform for a system with two excitonic sites and coupled vibrations ~\cite{GormanHaeffner18prx}, while a single non-Hermitian trapped-ion qubit has recently been demonstrated in experiments~\cite{DingZhang21prl,WangJingChen21pra,quinn2023observing}. The non-Hermitian VAET setup proposed in this work can be realized by combining elements of these prior experiments.  
Specifically, the energy sites of the chromophore donor-acceptor dimeric system shown in Figs.~\ref{fig:schematic}(a) and \ref{fig:schematic}(b) can be encoded in the internal electronic state of the ions. The tunneling coupling ($J$) can be realized and adjusted by a coherent laser drive, while the non-Hermitian terms ($\pm i\gamma$) can be realized and fine-tuned by manipulating gains and losses of the electronic levels by exchange with sources or sinks provided by additional laser drives [see Fig.~\ref{fig:schematic}(b)]. 

With laser heating, ions can absorb energy from additional laser beams with carefully chosen parameters, gaining kinetic energy. Such heating processes act as a source, supplying energy to the ions and increasing their motion. Using laser cooling, ions can be made to lose energy to the surrounding electromagnetic field by carefully tuning the frequency and intensity of the cooling lasers. Such cooling processes effectively act as a sink, removing energy from the ions and providing a damping effect on their motion. 

The excitonic interaction between the donor and acceptor, quantified by $\alpha$, can be engineered and adjusted by using a bichromatic laser beam aligned along the axis of the trap, while the donor-vibration coupling, quantified by $\kappa$, can be achieved and tuned through the design of two tones of a tightly focused laser beam localized on %each 
the donor ion. 
In trapped-ion quantum simulator experiments, the vibrational frequency $\nu$ is defined as the difference between the ion-crystal rocking-mode frequency and the frequency splitting between two tones of a bichromatic laser beam localized to the donor ion~\cite{GormanHaeffner18prx}. This definition allows for effective variation of the vibrational frequency in our setup. 
In addition to this effective tuning, trapped-ion quantum systems in general allow for the control of vibrational frequencies through the application of electromagnetic fields to the trapped ions, inducing specific oscillation frequencies. The vibrational frequency $\nu$ can be adjusted by modifying the frequency of the dynamic radiofrequency (RF) field, which can be accomplished by varying the amplitude and/or frequency of the RF signal directed into the trap~\cite{LeibfriedBlattMonroeWineland03rmp,Haffner08physrep,Bruzewicz19apr}. 
We note that, in contrast to the vibrational frequency, the energy gap $\Delta$ is more challenging to adjust~\cite{GormanHaeffner18prx}.

The non-Hermiticity of the dimer model in Eq.~(\ref{eq:H}) is inherited from the donor component, i.e., $H_d=J\sigma_x^{(d)}  -i\gamma \sigma_z^{(d)}$, which respects the ${\cal PT}$-symmetry, i.e., ${\cal PT}H_{d}{\cal PT} = H_{d}$ with parity operator ${\cal P}=\sigma_x^{(d)}$ and time-reversal operator ${\cal T}$ being a complex conjugation operation. The non-Hermitian donor system can therefore host an EP, as described in detail in Appendix~\ref{app:donor}. 
For the dimer consisting of a non-Hermitian donor and a coupled Hermitian acceptor, we can introduce an expanded ${\cal PT}$-symmetry operator, namely ${\cal P'T}$ with ${\cal P'}$ defined as ${\cal P}'=\sigma_x^{(d)}\otimes {\cal I}^{(a)}$. 
In the absence of coupling to vibration, i.e., $\kappa=0$, it is straightforward to show that ${\cal P'T}H (\kappa=0){\cal P'T} = H (\kappa=0)$ with ${\cal P}'=\sigma_x^{(d)}\otimes {\cal I}^{(a)}$, implying that the ${\cal PT}$-symmetry of the donor transfers to the dimer without vibration via this expanded operator. 

The requirements for VAET can be analyzed by consideration of the energetics of the uncoupled dimer, which are given in Appendix~\ref{sec:uphill}. The primary requirement for VAET in our dimer is that $\Delta >0$.  We shall focus on the uphill energy transfer processes, specifically the excitation transfer from the donor to the acceptor, as illustrated in Fig.~\ref{fig:donoreEnergylevels} of Appendix~\ref{sec:uphill}, which occur when $\Delta-J > \alpha/2$. 

We note that realizing such a non-Hermitian ${\cal PT}$-symmetric donor with a balanced distribution of gain and loss ($\pm \gamma$)  
can be challenging to achieve in experiments. However, an alternative approach can be adopted by constructing a passive ${\cal PT}$-symmetric system (possessing only loss, without gain terms) and relating this to a ${\cal PT}$-symmetric system by a loss offset.
For example, the passive ${\cal PT}$-symmetric Hamiltonian $\tilde{H}_d  = -i2\gamma\sigma^{(d)}_{+}\sigma^{(d)}_{-} + J\sigma_x^{(d)}$ is equivalent to $\tilde{H}_d = -i\gamma I + H_d $ where $-i\gamma I$ represents a loss offset and $H_d$ is the ${\cal PT}$-symmetric donor with balanced gain and loss. 

A mapping between the Pauli-operator-based model, e.g., Eq.~(\ref{eq:H}) for quantum simulation with trapped ions, and the widely-used Hamiltonian models for excitonic states of light harvesting systems has previously been established~\cite{LiSarovarWhaley22njp}. 
The effective Hamiltonian in the single electronic excitation subspace relevant to excitonic energy transfer can be expressed as $\tilde{H} =\Xi H\Xi= (-\Delta-i\gamma)\tilde{\sigma}_z + \alpha\tilde{\sigma}_x + \kappa \tilde{\sigma}_z(a+a^{\dagger}) + \nu a^{\dagger}a $, where $\Xi=|eg\rangle\langle eg| + |ge\rangle\langle ge|$, $\tilde{\sigma}_z=|eg\rangle\langle eg| - |ge\rangle\langle ge|$ and $\tilde{\sigma}_x=|eg\rangle\langle ge| + |ge\rangle\langle eg|$ with $|eg\rangle = |e\rangle_d\otimes|g\rangle_a$, $|ge\rangle = |g\rangle_d\otimes|e\rangle_a$, $\langle eg| = (|eg\rangle)^{\dagger}$, and $\langle ge| = (|ge\rangle)^{\dagger}$.  
In the absence of vibration, the effective two-level non-Hermitian system within the single excitation subspace exhibits eigenvalues given by $\pm\sqrt{\alpha^2-(\gamma-i\Delta)^2}$ and still possesses an exceptional point at $\gamma=\alpha$ when $\Delta=0$. However, this particular parameter regime does not favor the uphill energy transfer (i.e., the excitation transfer from the donor to the acceptor) assisted by vibrational modes that is the focus of the current work and we do not consider the single excitation subspace effective Hamiltonian any more in this work.

To demonstrate the effect of the non-Hermiticity on the VAET processes, we focus on the population transferred from the donor to the acceptor, which is quantified by the population of the state $|ge\rangle$ ($=|g\rangle_d\otimes|e\rangle_a$ representing the state in which the donor is at its ground state and the acceptor is at its excited state). 
All calculations use the initial condition $|eg\rangle$ ($=|e\rangle_d\otimes|g\rangle_a$ representing the state where donor is at its excited state and the acceptor is at its ground state), which can be reached optically from the ground state of the dimer by a $\pi/2$ pulse. 
This population is denoted by $P_a(t)$ and can be calculated for the dimer as
\begin{eqnarray}
P_a(t)&=&{\rm Tr}[\rho(t)|ge\rangle\langle ge|] ,
\label{eq:Pa_t}
\end{eqnarray}
where $\rho(t)$ represents the total density matrix. 
We also define the transfer efficiency, denoted by $\bar{P}_a$, as the average population accumulation over a given time period $t_f$, given by
\begin{eqnarray}
\bar{P}_a=\frac{1}{t_f}\int_0^{t_f}P_a(t)dt	.
\label{eq:bar_Pa}
\end{eqnarray}
The density matrix $\rho(t)$ is obtained as
\begin{eqnarray}
\rho(t)=\frac{U^{\dagger}\rho_{\rm dim}(0)\rho_v U}{{\rm Tr}[U^{\dagger}\rho_{\rm dim}(0)\rho_v U]} ,
\label{eq:rho_total}
\end{eqnarray}
where $\rho_{\rm dim}(0)=|eg\rangle\langle eg|$ is the initial state of the dimer (i.e., an excitation initially on the donor) and $U=e^{-iHt}$ with $H$ given in Eq.~(\ref{eq:H}) is the nonunitary time evolution operator. 
For the vibration, we consider a thermal initial state described by $\rho_v=e^{-\beta \nu a^{\dagger}a}/{\rm Tr}(e^{-\beta \nu a^{\dagger}a})$ 
with $\beta=1/k_{\rm B}T$, where $k_B$ denotes the Boltzmann constant and $T$ represents the temperature. This state is characterized by $k_BT$ and is relatively easy to prepare in trapped-ion quantum simulator experiments~\cite{GormanHaeffner18prx}. 
Throughout this work, we choose the Fock space size of the vibration as $N=50$, which is much larger than the average phonon number. This choice ensures the accuracy and convergence of the results. 

In the subsequent numerical calculations, specific parameter values have been adopted to provide a meaningful context for our analysis. Unless explicitly specified, we have chosen typical values within the domain of trapped-ion energy scales, namely $\alpha/J=1$, $\Delta/J=8$, $\nu/J=16.12$, $\kappa/J=0.3$, $k_B T/J=40$, and $t_f=22.5/J$. These values are motivated by the characteristic kHz frequencies observed in trapped-ion systems. For example, $\{J, \,\alpha, \,\Delta, \,\nu, \,\kappa, \,k_B T\} = 2\pi\times \{2.6, \,2.6, \,20.8, \,41.912, \,0.78, \,104\}$ kHz, and $t_f=1.378$ ms~\cite{GormanHaeffner18prx}. The value of $\Delta$ ($>J+\alpha/2$) ensures the uphill transfer of excitations from the donor to the acceptor, as detailed in Appendix~\ref{sec:uphill}. The vibrational frequency $\nu$, nearly resonant with the dimer transition, is determined by the dimeric energy structure ($J$, $\alpha$, $\Delta$, and the gain-loss variable $\gamma$). The chosen value of $\kappa$ places the system in the weak dimer-vibration coupling regime, and the vibrational temperature $k_BT$ is set high to ensure that the vibrational mode provides the necessary phonons to facilitate excitation energy transfer in the dimer. The robustness to different values of $\kappa$ and $k_BT$ is discussed in Sec.~\ref{sec:robustness}. It is important to note that alternative choices for the final evolution time $t_f$ do not affect the primary result of the non-Hermitian VAET spectral features, as discussed in Sec.~\ref{spectra}.

\section{EPs and dynamics in the absence of vibrational modes}
\label{sec:dimer}

\subsection{Exceptional points}

When the non-Hermitian chromophore dimer is decoupled from the vibrational mode ($\kappa=0$), its Hamiltonian is given simply by
\begin{eqnarray}
H_{\rm dim} &=&  -i\gamma\sigma_z^{(d)} + J\sigma_x^{(d)} + \Delta\sigma_z^{(a)} + \alpha\sigma_x^{(d)}\sigma_x^{(a)} 
\label{eq:H_dimer}
\end{eqnarray}
with eigenenergies $\lambda_j$ ($j=1,2,3,4$) given by 
\begin{eqnarray}
\lambda_1 &=&-\lambda_2 = -\sqrt{\xi - 2\sqrt{\alpha^2J^2+(J-\gamma)(J+\gamma)\Delta^2}} , \label{eq:eval_12}\\
\lambda_3 &=& -\lambda_4 = -\sqrt{\xi + 2\sqrt{\alpha^2J^2+(J-\gamma)(J+\gamma)\Delta^2}} , \label{eq:eval_34}
\end{eqnarray}
where $\xi=\alpha^2+J^2-\gamma^2+\Delta^2$. The eigenenergies are shown in Fig.~\ref{fig:schematic}(a). The corresponding eigenstates are denoted as $|\psi_i\rangle$.  
This chromophore dimer can exhibit two-fold degenerate second-order EPs, as we discuss in detail below. In this work we shall focus on the second-order EPs found for the parameter ratio  $\Delta/J=8$, analyzing the dynamics in particular for the case where the ratio of excitonic coupling to donor chromophore tunneling is unity, i.e., $\alpha/J=1$. 

\begin{figure}%[hbt!]
\centering
  \includegraphics[width=.99\columnwidth]{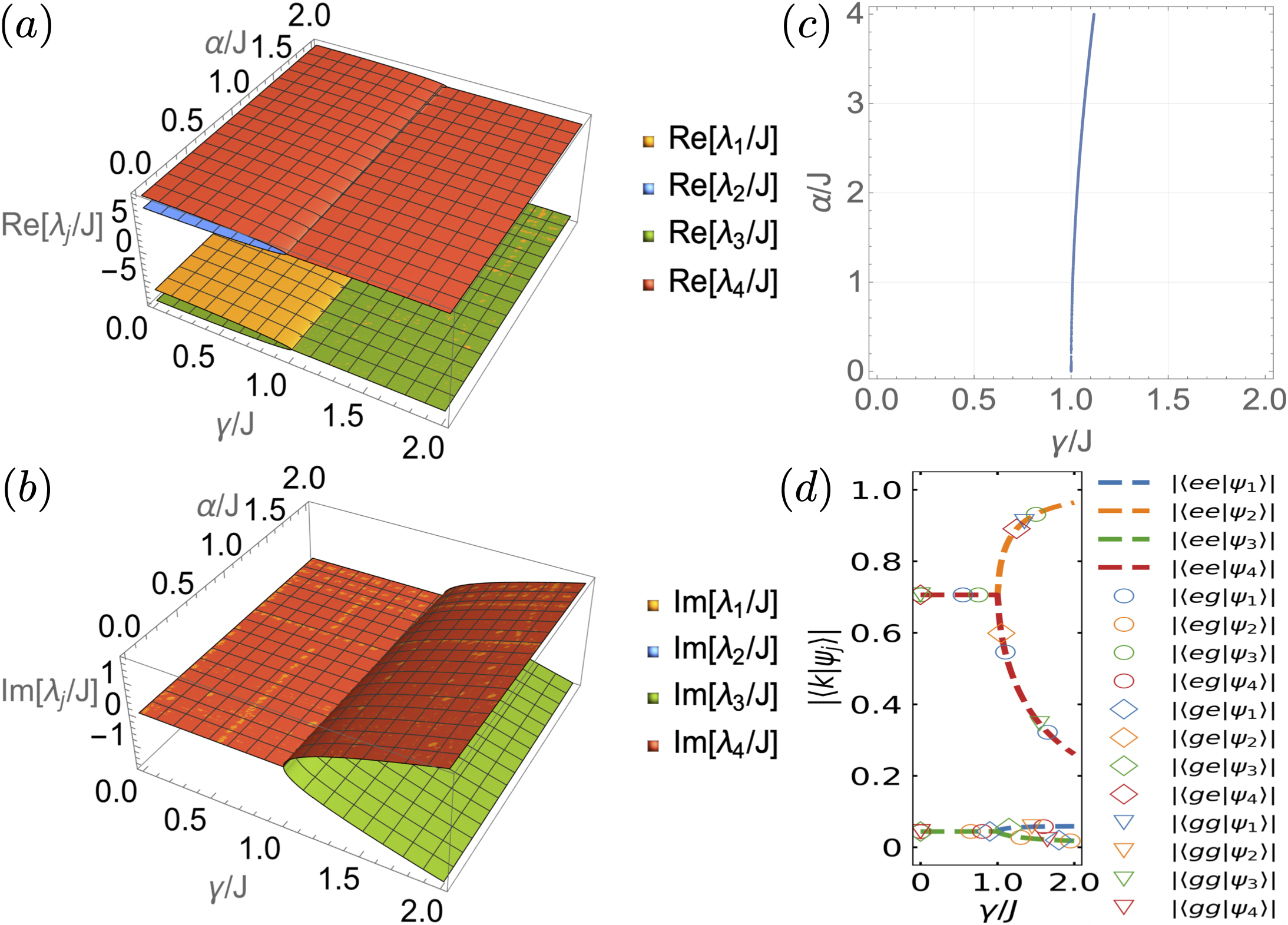}
\caption{(color online) (a, b) Real (a) and imaginary (b) parts of eigenenergies $\lambda_i$ for the non-Hermitian chromophore dimer $H_{\rm dim}$ with $\Delta/J=8$ and variable donor-acceptor coupling $\alpha$. Note the occurrence of two pairs of doubly degenerate energies along the line starting at $\gamma/J=1$ for $\alpha=0$. (c) Exceptional line composed of two pairs of degenerate eigenvalues for the non-Hermitian chromophore dimer $H_{\rm dim}$ with $\Delta/J=8$. The limit $\alpha =0$ represents the uncoupled dimer which contains the non-Hermitian donor monomer. (d) Projections at $\alpha/J =1$ of eigenstates $|\psi_j\rangle$ [i.e, $|\psi_1\rangle$ (blue), $|\psi_2\rangle$ (orange), $|\psi_3\rangle$ (green), and $|\psi_4\rangle$ (red)] onto the four basis states $|ee\rangle$ (dashed lines), $|eg\rangle$ (circles), $|ge\rangle$ (diamonds), and $|gg\rangle$ (down-pointing triangles). It is evident that there is a simultaneous coalescence of eigenvectors $|\psi_1\rangle$ and $|\psi_3\rangle$, and of eigenvectors $|\psi_2\rangle$ and $|\psi_4\rangle$, with each pair corresponding to a different degenerate eigenvalue [panel (a)], resulting in a two-fold degenerate second-order EP. Unless otherwise specified, all plots are made with parameters $\alpha/J=1$, $\Delta/J=8$, and $\kappa/J=0$.}
\label{fig:dimer_nonHermitian}
\end{figure}

Despite the fact that $H_{\rm dim}$ in Eq.~(\ref{eq:H_dimer}) is not Hermitian (i.e.,  $H_{\rm dim}\neq  H_{\rm dim}^{\dagger}$), its eigenenergies can still be real, since it is ${\cal PT}$-symmetric as noted above. 
This is evidenced in Figs.~\ref{fig:dimer_nonHermitian}(a) and \ref{fig:dimer_nonHermitian}(b), which display respectively the real and imaginary parts of the eigenenergies $\lambda_j$, as functions of the gain-loss rate $\gamma$ and the donor-acceptor coupling $\alpha$, for $\Delta/J=8$. 
It is evident that there are two lines of second-order exceptional points, one representing the simultaneous degeneracy of eigenenergies $\lambda_1$ and $\lambda_3$, and the other representing the degeneracy of eigenenergies $\lambda_2$ and $\lambda_4$. Remarkably, while the real component of the eigenenergies differ on the two lines, in the $(\gamma/J, \alpha/J)$ plane the lines are identical, resulting in a single, two-fold degenerate, line of second-order exceptional points in the parameter space $\{\gamma/J, \alpha/J\}$. 
This line is shown explicitly in Fig.~\ref{fig:dimer_nonHermitian}(c) to more clearly reveal how the second-order EP of the non-Hermitian monomer donor chromophore, which is located at $(\gamma/J,\alpha/J)=(1,0)$ (see Appendix~\ref{app:donor}), is smoothly transformed to the degenerate pair of second-order EPs of the dimer as the donor-acceptor (excitonic) coupling to the acceptor chromophore, $\alpha$, is turned on. These EPs of the coupled dimer are only slightly shifted, to larger values of the gain-loss parameter,  
as $\alpha/J$ is increased. Thus the dimer EP is very close to the donor monomer EP. 
Physically, this indicates that for the dimer EP, the excitonic coupling $\alpha$ plays a role similar to the internal donor tunneling $J$ in balancing the gain-loss rate $\gamma$.

We note that the second-order exceptional line in Fig.~\ref{fig:dimer_nonHermitian}(c) aligns closely with the vertical line $\gamma/J=1$. This is due to the parameter choice $\Delta/J=8$ which describes the uphill energy transfer of excitations from donor to acceptor, 
as discussed in the previous section and in Appendix~\ref{sec:uphill}. 
Corresponding exceptional lines for other values of $\Delta/J$, specifically for $\Delta/J=0$ and $\Delta/J=2$, are shown in Appendix~\ref{app:dimer}.

Figure~\ref{fig:dimer_nonHermitian}(d) shows the absolute value of the projection of the eigenstates $|\psi_j\rangle$ onto the four basis states $|k\rangle$ ($k=ee, eg, ge, gg$), for the parameter choice $\alpha/J=1$ and $\Delta/J = 8$. For each basis state, i.e., $|ee\rangle$ (dashed lines), $|eg\rangle$ (circles), $|ge\rangle$ (diamonds), or $|gg\rangle$ (down-pointing triangles), it is evident that the eigenstates of both pairs $|\psi_1\rangle$ (blue) and $|\psi_3\rangle$ (green), and  $|\psi_2\rangle$ (orange) and $|\psi_4\rangle$ (red) coalesce at a common second-order EP, indicating the two-fold degeneracy of this second-order EP.

\subsection{Non-Hermitian features of excitation energy transfer processes \label{sec:dimer:NHdynamics}}

The time evolution of the acceptor population, denoted as $P_a(t)$ [see Eq.~(\ref{eq:Pa_t})], 
is presented in Figs.~\ref{fig:dimer_nonHermitian_dynamics}(a) and \ref{fig:dimer_nonHermitian_dynamics}(b). These correspond respectively to the dynamics in ${\cal PT}$-symmetry unbroken phase [panel (a), with real eigenenergies shown in Figs.~\ref{fig:dimer_nonHermitian}(a) and \ref{fig:dimer_nonHermitian}(b)], and to the dynamics in the broken phase [panel (b), with complex eigenenergies].  
The two phases are separated by an EP, which for the case of $\alpha/J=1$ and $\Delta/J=8$ is located at $\gamma/J=1.00778\sim1$.

\subsubsection{$\gamma=0$}

\begin{figure}%[hbt!]
\centering
  \includegraphics[width=0.99\columnwidth]{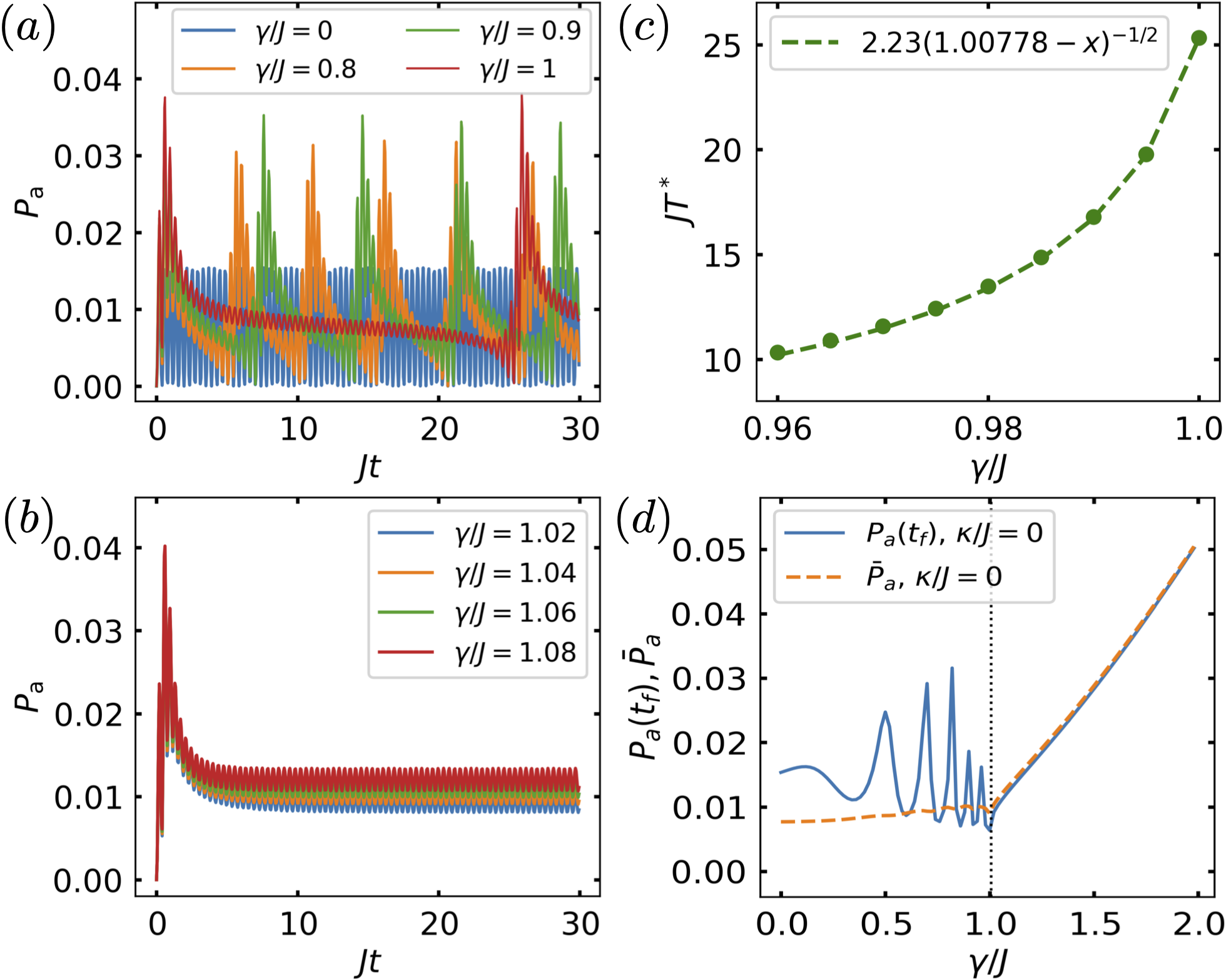}
\caption{(color online) 
(a, b) Non-Hermitian dynamics of $P_a(t)$ in ${\cal PT}$-symmetry unbroken (a) and broken (b) phases. (c) Dependence of the period of slow oscillations in panel (a), $T^*=2\pi/\lambda_{42(13)}$, on the gain-loss rate $\gamma$ when approaching the EP in the unbroken phase. The green dashed curve is an eye guide calculated from $y=2.23(1.00778-x)^{-1/2}$. (d) Dependence of $P_a(t_f)$ (blue solid curve) and ${\bar P}_a$ (orange dashed curve) on $\gamma/J$, showing the ${\cal PT}$-symmetry phase transition at $\gamma/J = 1$. Unless otherwise specified, all plots are made with parameters $\alpha/J=1$, $\Delta/J=8$, $\kappa/J=0$, and $t_f=22.5/J$. 
The initial state is $|eg\rangle$ in all calculations.}
\label{fig:dimer_nonHermitian_dynamics}
\end{figure}

We first analyze the Hermitian case $\gamma=0$ [blue curve in Fig.~\ref{fig:dimer_nonHermitian_dynamics}(a)]. In this case, $P_a(t)$ exhibits perfect Rabi oscillations with a period close to $0.39/J$ (twenty five oscillations in $Jt\le10$), corresponding to transitions between eigenstates $|\psi_{1(3)}\rangle$ and $|\psi_{4(2)}\rangle$ with a frequency $\lambda_{41(23)}\equiv\lambda_{4(2)}-\lambda_{1(3)}\approx 16.12J$ %16.124J 
(see the cyan up-down arrows in Fig.~\ref{fig:schematic}(a) and eigenenergies $\lambda_j$ at $\gamma=0$ in Appendix~\ref{app:dimer}). 
This fast oscillation of the population $P_a(t)$ at the acceptor (in state $|ge\rangle$) results from the donor-acceptor coupling, represented by $H_{\alpha}\equiv\alpha\sigma_x^{(d)}\sigma_x^{(a)}$ in Eq.~(\ref{eq:H_dimer}), which transfers the initial excitation from the donor state ($|eg\rangle$) to the acceptor state.  
The eigenstates of the decoupled Hamiltonian $\tilde{H}_{0}\equiv H_{\rm dim}^{\alpha,\gamma\rightarrow0}=J\sigma_x^{(d)}  +\Delta\sigma_z^{(a)}$, are  
\begin{eqnarray}
|\tilde{\psi}_{1(3)}\rangle\equiv|\psi_{1(3)}\rangle_{\alpha,\gamma\rightarrow0} &=& \frac{|e\rangle_{d}\pm|g\rangle_{d}}{\sqrt{2}}\otimes|g\rangle_{a} , \label{eq:psi_13_limit} \\
|\tilde{\psi}_{4(2)}\rangle\equiv|\psi_{4(2)}\rangle_{\alpha,\gamma\rightarrow0} &=& \frac{|e\rangle_{d}\pm|g\rangle_{d}}{\sqrt{2}}\otimes|e\rangle_{a} , \label{eq:psi_42_limit}
\end{eqnarray}
with corresponding eigenenergies $\lambda_{1(3)}=\pm J-\Delta$ and $\lambda_{4(2)}=\pm J+\Delta$, respectively,  
illustrated by grey dashed lines in Fig.~\ref{fig:schematic}(a). 
In the interaction picture with respect to $\tilde{H}_0$, the donor-acceptor coupling becomes 
\begin{eqnarray}
e^{i\tilde{H}_0t}H_{\alpha}e^{-i\tilde{H}_0} 
&=& \sum_{j,k} e^{i(\lambda_j-\lambda_k)t} \langle\tilde{\psi}_j|H_{\alpha}|\tilde{\psi}_k\rangle |\tilde{\psi}_j\rangle \langle\tilde{\psi}_k | \notag\\
&=& \alpha \big(e^{i\lambda_{41}t} |\tilde{\psi}_4\rangle \langle\tilde{\psi}_1 | + e^{i\lambda_{23}t} |\tilde{\psi}_2\rangle \langle\tilde{\psi}_3 |\big) , 
\end{eqnarray}
indicating that $H_{\alpha}$ is responsible for the fast oscillations with the oscillating period $2\pi/\lambda_{41(23)}$ observed in the blue cure of Fig.~\ref{fig:dimer_nonHermitian_dynamics}(a). 
Alternatively, we can express the initial and target states as superpositions of the eigenstates [i.e., $|eg\rangle=(|\psi_{1}\rangle_{\alpha,\gamma\rightarrow0} + |\psi_{3}\rangle_{\alpha,\gamma\rightarrow0})/\sqrt{2}$ or $|ge\rangle=(|\psi_{4}\rangle_{\alpha,\gamma\rightarrow0} - |\psi_{2}\rangle_{\alpha,\gamma\rightarrow0})/\sqrt{2}$], and calculate the transition matrix element of the interaction $\langle ge|\alpha\sigma_x^{(d)}\sigma_x^{(a)}|eg\rangle$. This calculation shows that non-zero elements only exist between $|\psi_{1(3)}\rangle_{\alpha,\gamma\rightarrow0}$ and $|\psi_{4(2)}\rangle_{\alpha,\gamma\rightarrow0}$, corresponding to the fast oscillations induced by the donor-acceptor coupling $H_{\alpha}$.

\subsubsection{$\gamma\neq0$}

Upon entering the ${\cal PT}$-symmetry unbroken phase (i.e., $0<\gamma/J \lesssim 1$), slow oscillations of $P_a(t)$ emerge that are superimposed on the fast oscillation, with a reduced amplitude compared to the Hermitian case. These slow oscillations result in a longer oscillating period as the system approaches the EP, evidenced by, e.g., the orange ($\gamma/J=0.8$) or green ($\gamma/J=0.9$) curve in Fig.~\ref{fig:dimer_nonHermitian_dynamics}(a). 
These oscillations in the unbroken phase are induced by the non-Hermitian term $-i\gamma\sigma_z^{(d)}$ in Eq.~(\ref{eq:H_dimer}). They correspond to transitions between eigenstates $|\psi_{4(1)}\rangle$ and $|\psi_{2(3)}\rangle$, with a transition frequency $\lambda_{42(13)}\equiv\lambda_{4(1)}-\lambda_{2(3)}$ [see the orange up-down arrows in Fig.~\ref{fig:schematic}(a)]. For example, the transition frequency takes the value of $1.217J$ for $\gamma/J=0.8$, resulting in a period of $5.16/J$ (approximately two oscillations when $Jt\le10$) for the orange curve in Fig.~\ref{fig:dimer_nonHermitian_dynamics}(a). 
The longer oscillating period near the EP is due to the decrease of $\lambda_{13(42)}$ as $\gamma$ increases, as can be seen from Eqs.~(\ref{eq:eval_12}) and (\ref{eq:eval_34}) or Fig.~\ref{fig:dimer_nonHermitian}(a). 
These slow $\gamma$-induced oscillations can be understood by employing the eigenstates of $H_{\rm dim}^{\alpha,\gamma\rightarrow0}$ %$|\psi_1\rangle_{\alpha,\gamma\rightarrow 0}$ and $|\psi_3\rangle_{\alpha,\gamma\rightarrow 0}$ (or $|\psi_4\rangle_{\alpha,\gamma\rightarrow 0}$ and $|\psi_2\rangle_{\alpha,\gamma\rightarrow 0}$) 
in Eq.~(\ref{eq:psi_13_limit}) [or Eq.~(\ref{eq:psi_42_limit})] for the acceptor at its ground (or excited) state. Considering the initial state $|eg\rangle=(|\psi_{1}\rangle_{\alpha,\gamma\rightarrow0} + |\psi_{3}\rangle_{\alpha,\gamma\rightarrow0})/\sqrt{2}$ and the target state $|ge\rangle=(|\psi_{4}\rangle_{\alpha,\gamma\rightarrow0} - |\psi_{2}\rangle_{\alpha,\gamma\rightarrow0})/\sqrt{2}$ in terms of these eigenstates, the non-vanishing transition matrix elements of the non-Hermitian term $\langle ge| -i\gamma\sigma_z^{(d)} |eg\rangle$  
only appear between $|\psi_{1(4)}\rangle_{\alpha,\gamma\rightarrow0}$ and $|\psi_{3(2)}\rangle_{\alpha,\gamma\rightarrow0}$. This, together with Eq.~(\ref{eq:psi_13_limit}) [or Eq.~(\ref{eq:psi_42_limit})], implies oscillations with an oscillating period $2\pi/\lambda_{13(42)}$ between states $(|e\rangle_d +|g\rangle_d)/\sqrt{2}$ and $(|e\rangle_d -|g\rangle_d)/\sqrt{2}$ in the donor subspace.  
Moreover, during each period of the slow oscillation, there is a rapid growth of $P_a(t)$ followed by a slow decrease. This behavior can be understood as the gain-loss rate $\gamma$ more easily promoting the transfer of an initial excitation at the donor where it is added, than decreasing the excitation at the acceptor. 
The $\gamma$-induced slow oscillations between $|\psi_{1(4)}\rangle$ and $|\psi_{3(2)}\rangle$ distort the $\alpha$-induced fast oscillations between $|\psi_{1(3)}\rangle$ and $|\psi_{4(2)}\rangle$, giving rise to the modulated Rabi-like oscillations in the unbroken phase evident in Fig.~\ref{fig:dimer_nonHermitian_dynamics}(a).

In the ${\cal PT}$-symmetry broken phase (i.e., $\gamma/J\gtrsim1$), Fig.~\ref{fig:dimer_nonHermitian_dynamics}(b) %llustrates 
shows that the rapid growth of the acceptor population $P_a(t)$ is followed by approach to a non-equilibrium steady state which is characterized by small amplitude fast oscillations. 
The population in this steady state increases slightly with $\gamma/J$.  
Similar to the Hermitian case %in the unbroken symmetry phase 
[$\gamma=0$, blue curve in Fig.~\ref{fig:dimer_nonHermitian_dynamics}(a)], these fast oscillations correspond to transitions between $|\psi_{1(3)}\rangle$ and $|\psi_{4(2)}\rangle$. 
The non-equilibrium steady state arises primarily from the purely imaginary transition frequency between eigenstates $|\psi_{1(4)}\rangle$ and $|\psi_{3(2)}\rangle$. For example, the transition frequency is approximately $\lambda_{13(42)}=\lambda_{1(4)}-\lambda_{3(2)}\sim \{i0.312J, \, i0.51 J, \, i0.652J, \, i0.77J\}$ for $\gamma/J=\{1.02, \, 1.04, \, 1.06, \, 1.08\}$, respectively (see Appendix~\ref{app:dimer}). This steady state can be regarded as the symmetry-broken analog of the $\gamma$-induced slow oscillations observed in Fig.~\ref{fig:dimer_nonHermitian_dynamics}(a), but now with an infinite period resulting from the vanishing real part of the transition frequency in the symmetry-broken phase. 
From this perspective, the larger steady-state population for the larger $\gamma$ value evident in Fig.~\ref{fig:dimer_nonHermitian_dynamics}(b) can be attributed to the more rapid growth resulting from transfer of an initial excitation promoted by the gain-loss rate, as discussed in the preceding paragraph.

To conclude this section, in Figs.~\ref{fig:dimer_nonHermitian_dynamics}(c) and \ref{fig:dimer_nonHermitian_dynamics}(d) we present, respectively, the period $T^*$ of slow oscillations in the unbroken symmetry phase, and the behavior of $P_a(t_f)$ and $\bar{P}_a$ over the full range of $\gamma/J$. 
Figure~\ref{fig:dimer_nonHermitian_dynamics}(c) displays the dependence of $T^*=2\pi/\lambda_{42(13)}$, i.e., the slow oscillations for the $\lambda_{42}$ and $\lambda_{13}$ %42 and 13 
transitions of Fig.~\ref{fig:schematic}(a), on the gain-loss rate $\gamma$ close to the exceptional point. On decreasing $\gamma$ away from the EP in the unbroken phase, $T^*$ exhibits an inverse square-root relation, indicated by the green dashed curve as a visual guide. This dependence arises from the second-order nature of the EP. Specifically, the eigenenergies $\lambda_j$ as well as the resulting energy difference $\lambda_{42(13)}$ are proportional to the square root of a small deviation in $\gamma$ from the EP at $\gamma/J=1.00778\sim1$, leading to an inverse square root dependence of the period $T^*$. 
The green dashed curve also aligns with the observation that $T^*$ tends to infinity at the EP where $\lambda_{13}=\lambda_{42}\rightarrow0$.

Figure~\ref{fig:dimer_nonHermitian_dynamics}(d) shows that both the acceptor population $P_a(t=t_f)$ at a given time $t_f$, and the acceptor population accumulation $\bar{P}_a$ up to this time [Eq.~(\ref{eq:bar_Pa})], provide qualitative order parameters for the transition between the unbroken and broken ${\cal PT}$-symmetry phases. 
Apart from the expected difference between these two quantities in the regime $\gamma/J\lesssim1$ (i.e., with or without oscillating behaviors), it is evident that both $P_a(Jt_f=22.5)$ (blue solid curve) and $\bar{P}_a$ (orange dashed curve) undergo significant changes at the EP ($\gamma/J\sim1$), indicating the location of a ${\cal PT}$-symmetry phase transition.  
The monotonically increasing behavior of $P_a(t=t_f)$ or $\bar{P}_a$ for $\gamma/J\gtrsim1$ is attributed to the presence of the non-equilibrium steady state with a higher population for larger $\gamma$, that was observed in Fig.~\ref{fig:dimer_nonHermitian_dynamics}(d) and commented on above.

\section{Enhanced vibrationally assisted energy transfer near an EP \label{sec:VAET_spectrum}}

The relatively low population transported from the donor to the acceptor that is shown in Figs.~\ref{fig:dimer_nonHermitian_dynamics}(a), \ref{fig:dimer_nonHermitian_dynamics}(b), and \ref{fig:dimer_nonHermitian_dynamics}(d) can be increased by introducing coupling to vibrations ($\kappa\neq0$). 
We show in this section that the resulting excitation energy transfer process by absorption of phonons from a vibrational mode, which is the non-Hermitian extension of VAET in the Hermitian case ($\gamma=0$), can be significantly enhanced near an EP.  
After briefly summarizing the Hermitian VAET features, we shall explicitly analyze the efficiency of VAET as a function of the distance from the EP of the system without vibrations
that was characterized in the previous section. We first present the non-Hermitian dynamics of the entire system including the vibration, under the VAET resonance conditions, %  
which are $\nu=\lambda_{23}=\lambda_{41}=16.12J$ for one-phonon VAET and $2\nu=\lambda_{23}=\lambda_{41}=16.12J$ for two-phonon VAET [see Fig.~\ref{fig:schematic}(a)]. We then probe the spectral features of the VAET processes by continuously scanning the vibrational frequency $\nu$.  
Our calculations are based on analysis in the weak dimer-vibration coupling $\kappa$ regime, to avoid population of the vibronic states that characterize the strong coupling regime~\cite{LiSarovarWhaley21njp}. We also assume an uphill transfer ($\Delta-J>\alpha/2$), which we realize with the specific parameter choice $\Delta/J=8$ and $\alpha/J = 1$.

\begin{figure}%[hbt!]
\centering
  \includegraphics[width=.99\columnwidth]{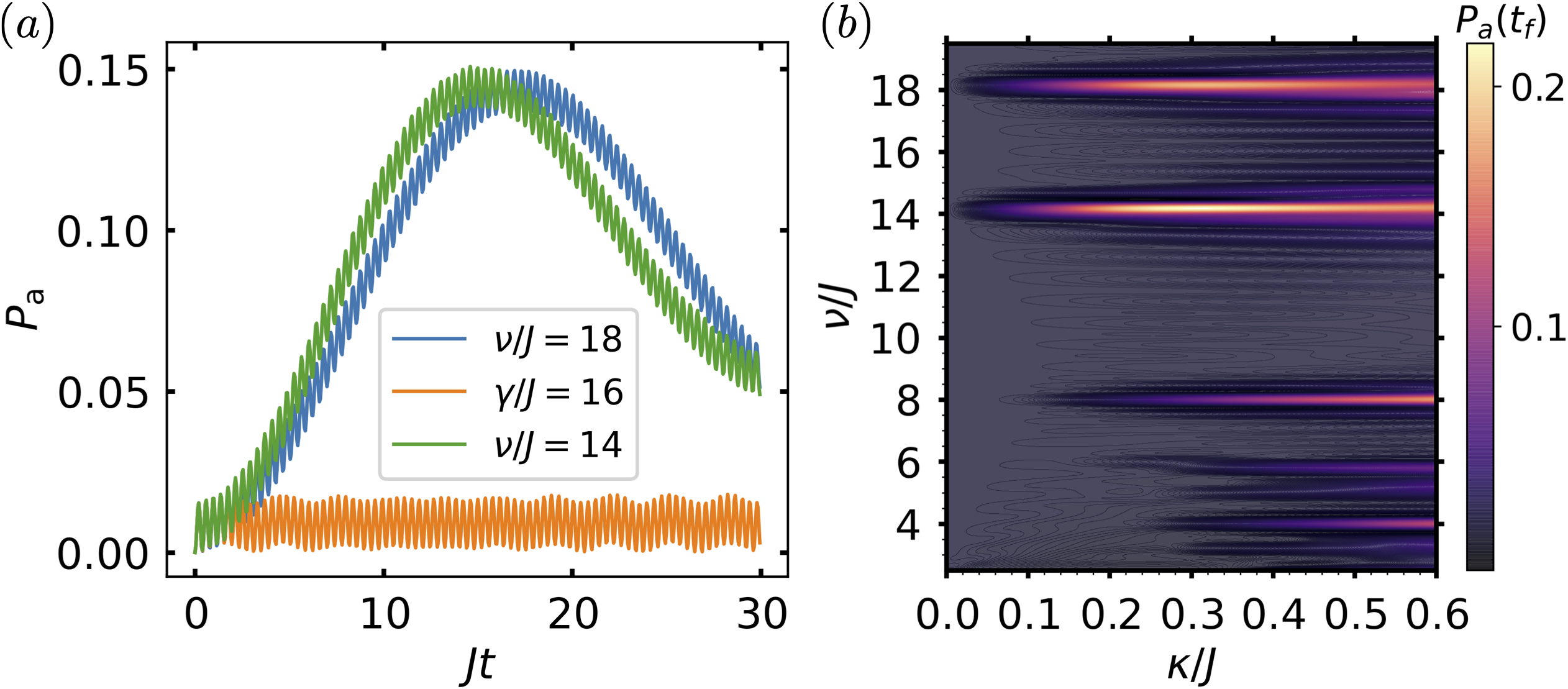}
\caption{(color online) (a) Dynamics of $P_a(t)$ for the one-phonon VAET in the Hermitian limit with $\nu/J=14, 16, 18$. (b) Hermitian spectra of the VAET processes as a function of $\kappa/J$ and $\nu/J$, represented by $P_a(t_f)$ with $t_f=22.5/J$. Unless otherwise specified, all plots are made with parameters $\gamma=0$, $\alpha/J=1$, $\Delta/J=8$, $\kappa/J=0.3$, and with the initial state $|eg\rangle$.}
\label{fig:HermitianVAET}
\end{figure}

\subsection{Hermitian VAET \label{sec:Hermitian VAET}}

Figure~\ref{fig:HermitianVAET} summarizes the features of VAET for the Hermitian system. The dynamics in panel (a) show that, for a specific dimer-vibration coupling of $\kappa/J=0.3$, the blue and green curves representing $\nu/J=18$ and $\nu/J=14$ respectively in $P_a(t)$, display higher population at the acceptor than both the blue curve ($\gamma=\kappa=0$) in Fig.~\ref{fig:dimer_nonHermitian_dynamics}(a) for the case of no vibration, and the orange curve depicting $\nu/J=16$, which is nearly resonant with the dimer transition but is not allowed by the phonon absorption mechanism (as explained below). 
This pair of VAET processes, representing transitions in the Hermitian limit between the eigenstates $|\psi_3\rangle$ and $|\psi_4\rangle$ or between $|\psi_1\rangle$ and $|\psi_2\rangle$, are depicted as blue and green up-down arrows in Fig.~\ref{fig:schematic}(a). These processes correspond to the absorption of a phonon from the vibrational modes $\nu=\lambda_{43}$ or $\lambda_{21}$. 
This observation implies that the VAET processes ($\kappa\neq0$) exhibit sensitivity to the chosen vibration frequency. This frequency dependence is further exemplified in panel (b) of Fig.~\ref{fig:HermitianVAET}, which summarizes the dependence of the population transfer $P_a(Jt_f=22.5)$ on the vibrational frequency $\nu$ and the vibrational coupling strength $\kappa$. 
Figure~\ref{fig:HermitianVAET}(b) shows that for a fixed vibration frequency, such as $\nu/J=18$ or $\nu/J=14$, the population transfer initially increases with $\kappa$, followed by a subsequent, albeit very slight, suppression as $\kappa$ is further increased to the extent that it causes a departure from the weak dimer-vibration coupling regime and leads to the formation of vibronic states with strong mixing of excitonic and vibrational degrees of freedom~\cite{LiSarovarWhaley21njp}.

\begin{figure}%[hbt!]
\centering
  \includegraphics[width=.99\columnwidth]{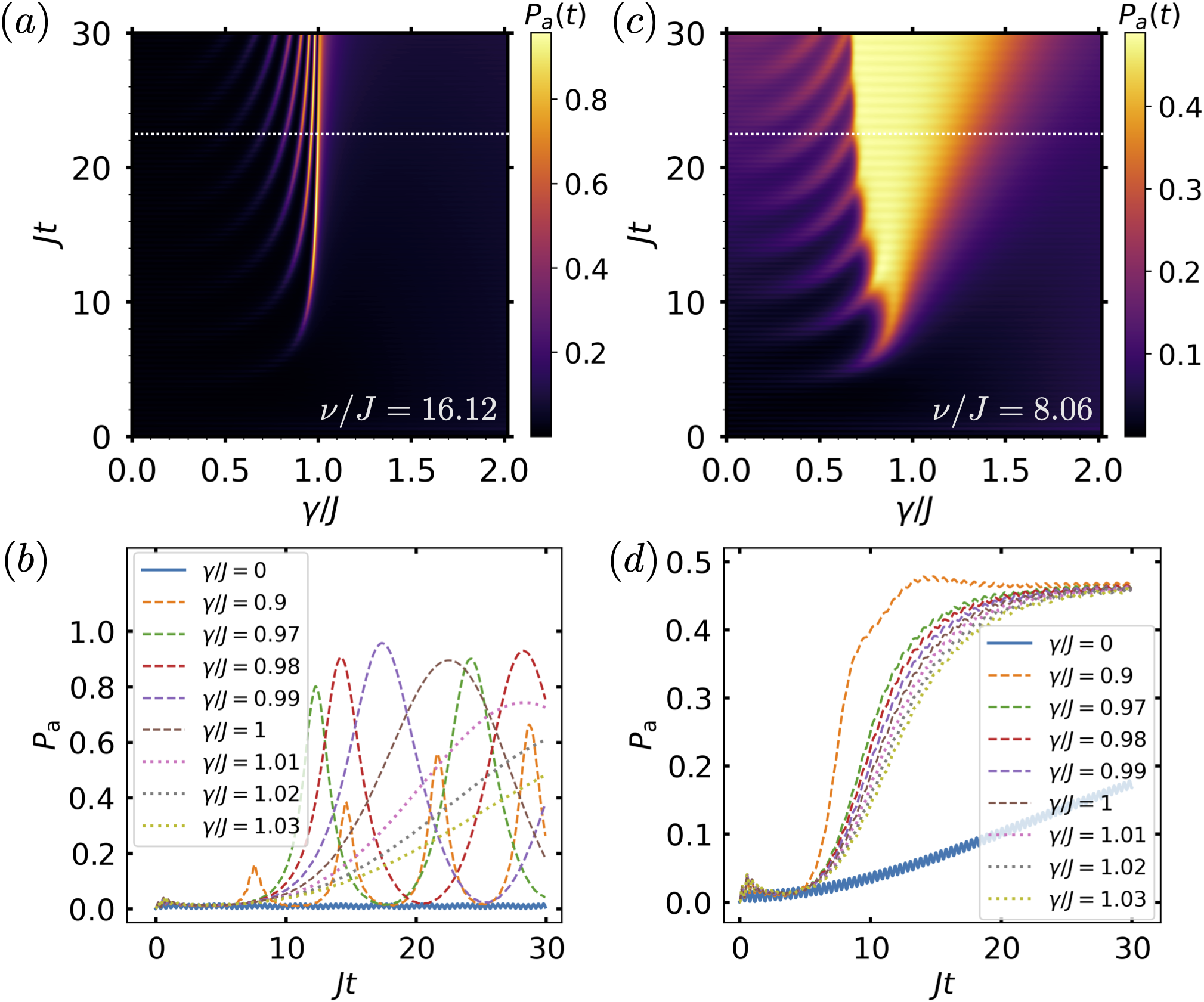}
\caption{(color online) (a, b) 
Non-Hermitian dynamics represented by $P_a(t)$, shown as a function of time $t$ for several values of gain-loss parameter $\gamma$ for one-phonon VAET processes, using $\nu/J=16.12$. 
(c, d)  $P_a(t)$ versus time $t$ for several values of gain-loss parameter $\gamma$ for two-phonon VAET processes, using $\nu/J=8.06$. The white dotted line in panels (a) and (c) represents $Jt=22.5$. Panels (b) and (d) show $P_a(t)$ for nine specific values of $\gamma$, starting with $\gamma=0$ (Hermitian case) and eight values for non-Hermitian systems, with $\gamma/J=0.9, 0.97, 0.98, 0.99, 1, 1.01, 1.02, 1.03$. 
Unless otherwise specified, all plots were obtained with the parameter values $\alpha/J=1$ and $\Delta/J=8$, for which a two-fold degenerate second-order EP is located at $\gamma/J=1.00778\sim1$, and 
$\kappa/J=0.3$, $k_BT/J=40$. The initial state is $|eg\rangle$ in all calculations.}
\label{fig:vaet_nonHermitian_dynamics}
\end{figure}

\subsection{Dynamical features of non-Hermitian VAET \label{sec:VAET_dynamics}} 

The non-Hermitian dynamics of the one-phonon VAET processes, quantified by the acceptor population $P_a(t)$, are shown in  Fig.~\ref{fig:vaet_nonHermitian_dynamics}(a, b) (with $\nu/J=16.12$), while the dynamics of the two-phonon VAET processes are shown in Fig.~\ref{fig:vaet_nonHermitian_dynamics}(c, d) (with $\nu/J=8.06$). 
Figures~\ref{fig:vaet_nonHermitian_dynamics}(b) and \ref{fig:vaet_nonHermitian_dynamics}(d) show the dynamics for nine specific values of the gain-loss parameter $\gamma$, extracted from Figs.~\ref{fig:vaet_nonHermitian_dynamics}(a) and \ref{fig:vaet_nonHermitian_dynamics}(c), respectively.

\subsubsection{One-phonon VAET dynamics}

Figure~\ref{fig:vaet_nonHermitian_dynamics}(a) shows that the acceptor population is initially low in the Hermitian limit ($\gamma=0$) and becomes more pronounced on approaching the EP (located at $\gamma/J=1.00778\sim1$ when $\alpha/J=1$ and $\Delta/J=8$) from the unbroken phase.  
The gradual increase in population as $\gamma$ moves toward the EP indicates that the one-phonon VAET processes (with $\nu=\lambda_{23}=\lambda_{41}=16.12J$) can be enhanced in the presence of non-Hermiticity. 
As $\gamma$ is further increased and the system enters the broken symmetry phase, $P_a(t)$ is suppressed again, in contrast to the non-equilibrium steady state in the absence of vibrations, 
where a higher population is observed for larger $\gamma$ [see Fig.~\ref{fig:dimer_nonHermitian_dynamics}(b)]. Therefore, there is also an enhancement of the VAET processes as $\gamma$ approaches the EP from the broken phase.
We also see that for a fixed time, i.e., a fixed $Jt$ value, the enhancement on approaching the EP from the unbroken symmetry phase is accompanied by oscillatory behavior with respect to $\gamma$. This is further illustrated by the comparison of $P_a(t)$ over the nine specific values of $\gamma$ that is shown in Fig.~\ref{fig:vaet_nonHermitian_dynamics}(b). For instance, the first peak of $P_a(t)$ reaches $0.957$ at $Jt=17.37$ for $\gamma/J=0.99$ (purple curve) or $0.895 $ at $Jt=22.52$ for $\gamma/J=1$ (brown curve). These values are at least forty times greater than the peak value observed in the Hermitian case, see Fig.~\ref{fig:vaet_nonHermitian_dynamics}(b), which shows a peak value of $\sim 0.02$ (represented by the blue curve with $\gamma=0$). We refer the reader also to Fig.~\ref{fig:HermitianVAET}(a) which exhibits similar peak values in the orange curve but with a slightly different vibrational frequency $\nu/J=16$. 
Interestingly, we note that the peak population for $\gamma/J=0.99$ is slightly higher than that for $\gamma/J=1$. This difference is attributed to a slight shift away from the EP position $\gamma/J=1.00778\sim1$ for the coupling value $\kappa/J=0.3$. To support this interpretation, an additional calculation was made with $\kappa/J=0.1$, which is closer to the EP at $\gamma/J=1.00778\sim1$. This calculation now explicitly shows that the peak population becomes higher, the closer the gain-loss parameter $\gamma$ is to its value at the EP. 
We show below that this enhancement of the VAET processes is a result of phonon absorption, resulting from the donor-vibration interaction in Eq.~(\ref{eq:H}) being maximally favorable at the two-fold degenerate second-order EP, allowing four simultaneous transitions between  
eigenstates associated with distinct EPs to be excited by a single phonon.

\subsubsection{Two-phonon VAET dynamics}

Figure~\ref{fig:vaet_nonHermitian_dynamics}(c) presents the two-phonon VAET processes. These also exhibit an %weaker 
enhancement relative to the $\gamma=0$ Hermitian case, but this is now weaker than the enhancement seen in the single-photon VAET of Fig.~\ref{fig:vaet_nonHermitian_dynamics}(a). In particular, for a given value of time $t$, i.e., fixed $Jt$ value, the oscillations with respect to $\gamma$ are now relatively far away from the EP and also saturate at values of $\gamma/J$ within the unbroken symmetry phase, i.e., they do not continue to increase on further approach to the EP. This early saturation of the enhancement before the EP is also evident on comparison of the curves in Fig.~\ref{fig:vaet_nonHermitian_dynamics}(d). We attribute this saturated enhancement to the coupling strength $\kappa$ that is relatively strong for the two-phonon process with $\nu/J=8.06$ and thereby enables the dimer-vibration system to arrive quickly to a steady state, compared to the one-phonon VAET case ($\nu/J=16.12$). To support this interpretation, we have conducted additional calculations with smaller values of $\kappa$ than used in  Figs.~\ref{fig:vaet_nonHermitian_dynamics}(c) and \ref{fig:vaet_nonHermitian_dynamics}(d). We observed two-phonon VAET dynamics similar to those in Fig.~\ref{fig:vaet_nonHermitian_dynamics}(a) or \ref{fig:vaet_nonHermitian_dynamics}(b), but with reduced populations. 
Our conclusion that the coupling strength becomes relatively strong for the two-phonon VAET process, with at the same time a smaller frequency than that of the one-phonon case, is further supported by a more pronounced modulation of Rabi-like oscillations in the Hermitian limit [blue curve in Fig.~\ref{fig:vaet_nonHermitian_dynamics}(d) with $\gamma=0$] than in the one-phonon case [blue curve in Fig.~\ref{fig:vaet_nonHermitian_dynamics}(b)] or the no-phonon case $\kappa=\gamma=0$ [blue curve with Rabi oscillations in Fig.~\ref{fig:dimer_nonHermitian_dynamics}(a)].

\subsection{Spectral features \label{spectra}}

In addition to the two specific values, $\nu/J=16.12$ and $8.06$, that we have used above for analysis of the one- and two-phonon VAET processes, respectively, it is possible to investigate the dynamics of energy transfer processes across the entire range of the non-Hermitian parameter space by continuously scanning the frequency of the vibrational mode, i.e., $\nu$. Figures~\ref{fig:vaet_nonHermitian_spectrum}(a) and \ref{fig:vaet_nonHermitian_spectrum}(b) illustrate the resulting spectra of VAET processes, represented by $P_a(t_f)$ and $\bar{P}_a$, respectively, as a function of $\gamma/J$ and $\nu/J$ for a weak phonon coupling value $\kappa/J=0.3$.

\subsubsection{Hermitian VAET spectrum}

\begin{figure}%[hbt!]
\centering
  \includegraphics[width=.99\columnwidth]{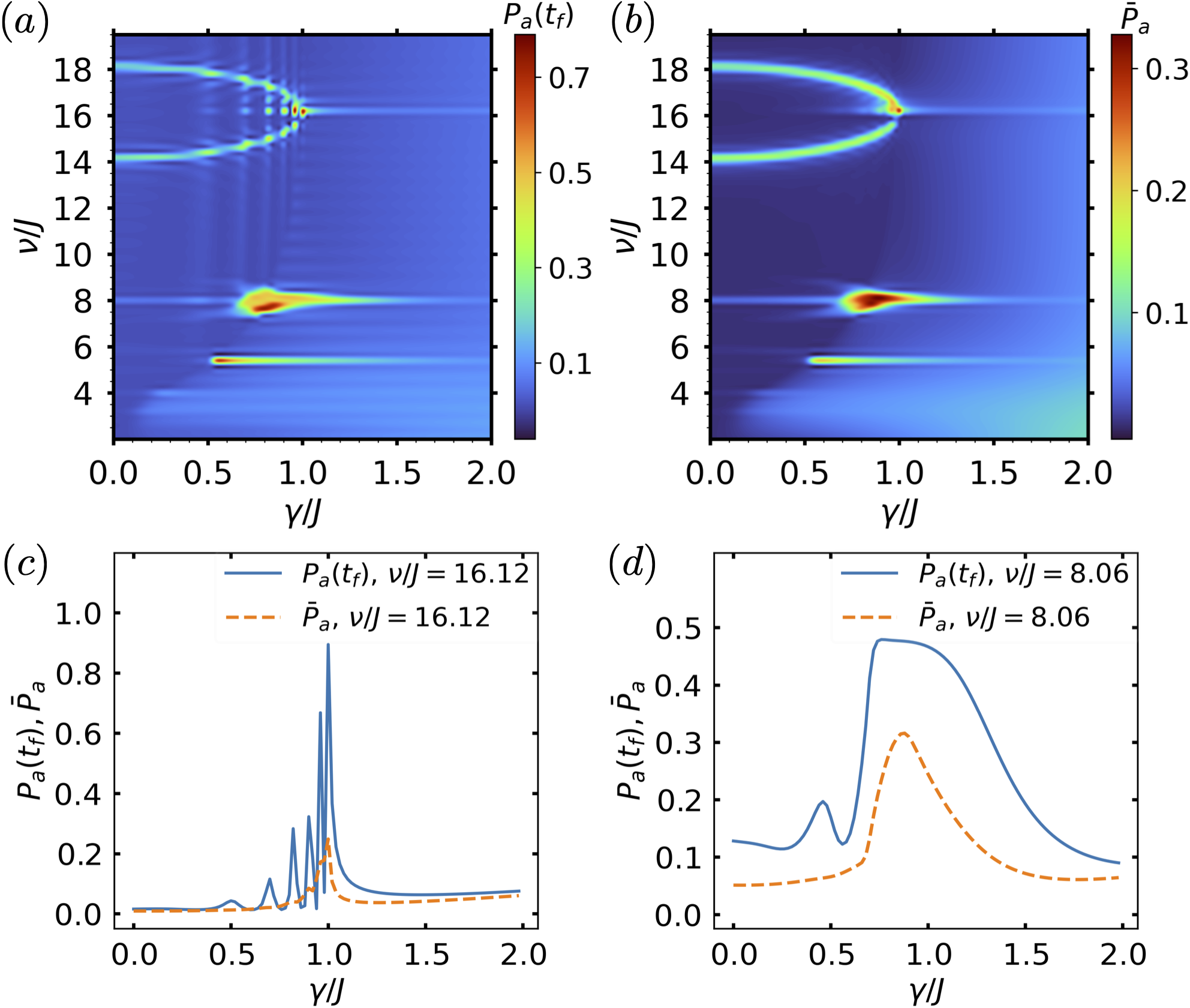}
\caption{(color online)
(a, b) Non-Hermitian spectra of the VAET processes as a function of $\gamma/J$ and $\nu/J$, represented by $P_a(t_f)$ in panel (a) and by $\bar{P}_a$ in panel (b). For these plots we take $t_f=22.5/J$, corresponding to the time at which $P_a$ is maximal for $\gamma/J=1$ according to the brown curve in Fig.~\ref{fig:vaet_nonHermitian_dynamics} (b). 
(c, d) One-dimensional spectra of VAET processes at $t_f = 22.5/J$, with $\nu/J=16.12$ (c) and $\nu/J=8.06$ (d), taken as cuts through the two-dimensional spectra in panels (a) and (b). Unless otherwise specified, all plots were obtained with the parameter values $\alpha/J=1$ and $\Delta/J=8$, for which two second-order EPs are located at $\gamma/J=1.00778\sim1$, and 
$\kappa/J=0.3$, $k_BT/J=40$. The initial state is $|eg\rangle$ in all calculations.}
\label{fig:vaet_nonHermitian_spectrum}
\end{figure}

In the Hermitian limit ($\gamma=0$), Fig.~\ref{fig:vaet_nonHermitian_spectrum}(a) shows one-phonon VAET processes occurring at $\nu/J\sim14$ and $18$, corresponding to transitions between eigenstates $|\psi_1\rangle$ and $|\psi_2\rangle$, or $|\psi_3\rangle$ and $|\psi_4\rangle$, respectively, that are accompanied by an absorption of a phonon from the vibrational mode $\nu=\lambda_{21}$ or $\lambda_{43}$. These one-phonon VAET processes, together with their corresponding dynamics displayed in Fig.~\ref{fig:HermitianVAET}(a), are indicated by green or blue up-down arrows in Fig.~\ref{fig:schematic}(a) and the eigenenergies $\lambda_i$ are given in Appendix~\ref{app:dimer}. 
%It is important to note that 
This VAET excitation energy transfer process is distinct from the fast oscillations between $|\psi_{1(3)}\rangle$ and $|\psi_{4(2)}\rangle$ with $\lambda_{41(23)}\sim16.12J$ [see the blue curve in Fig.~\ref{fig:dimer_nonHermitian_dynamics}(a) and cyan up-down arrows in Fig.~\ref{fig:schematic}(a)] that are observed in the absence of the vibrational mode.

\subsubsection{Non-Hermitian VAET spectrum}

Interestingly, as $\gamma$ increases from zero going into the non-Hermitian regime, the two one-phonon VAET processes 
originating from their respective Hermitian point at $\nu/J\sim18$ or $14$  
move closer to each other and eventually converge at $\nu/J=16.12$ at the EP (where $\gamma/J=1.00778\sim1$) [see Fig.~\ref{fig:vaet_nonHermitian_spectrum}(a)].
Moreover, at the EP, the population of the VAET process becomes more pronounced compared to both the population in the unbroken phase ($\gamma/J\lesssim1$) and in the broken phase ($\gamma/J\gtrsim1$) [see the white dotted %unmarked horizontal 
line $Jt=22.5$ in Fig.~\ref{fig:vaet_nonHermitian_dynamics}(a)]. 
This intriguing behavior of the two one-phonon VAET processes in response to changes in $\gamma/J$ arises from the coalescing of eigenstates and/or eigenenergies, as depicted in Figs.~\ref{fig:dimer_nonHermitian}(a, b, d), and is reflected in the shape of the maximal probabilities in the upper left quadrants of Figs.~\ref{fig:vaet_nonHermitian_spectrum}(a) and \ref{fig:vaet_nonHermitian_spectrum}(b). 

This phenomenon suggests a novel fluorescence-detected vibrational spectroscopy approach for probing both EPs and ${\cal PT}$-symmetry phase transitions in non-Hermitian quantum systems, namely by analysis of the spectrum of a weakly coupled vibrational mode. 
We note that the traditional approach to analysis of vibrational spectra considers the spectral amplitude at a given frequency/wavelength, which measures the intensity of phonons emitted at that specific frequency/wavelength. Higher spectral amplitudes indicate more intense emission, while lower spectral amplitudes correspond to weaker emission. 
We can take advantage of this in trapped-ion experiments when measuring the acceptor population by fluorescence detection~\cite{LeibfriedBlattMonroeWineland03rmp}, at a given vibrational frequency/wavelength. 
This population constitutes a measure of the energy transfer that is assisted by phonons emitted from the enabling vibrational mode. Higher acceptor population indicates a more intense emission of phonons, just as in the conventional vibrational spectrum analysis, thereby providing a measure of the vibrationally assisted energy transfer.  
Notably, the maximum acceptor population at the EP ($\gamma/J=1.00778\sim1$) that corresponds to the one-phonon VAET process with $\nu/J=16.12$ is attributed to the two-fold second-order non-Hermitian degeneracy at this point. 
This degeneracy results in four simultaneous transitions (with $\lambda_{41}=\lambda_{23}=\lambda_{21}=\lambda_{43}$ given that $\lambda_1=\lambda_3$ and $\lambda_2=\lambda_4$ at the two-fold degenerate second-order EP) that are excited by a single phonon. 
The phonon-absorption mechanism described in the next subsection below provides further support for this interpretation.

Other than single phonons, energy transfer processes involving the absorption of two or even three phonons from the coupled vibrational mode are also observed. These processes occur at $\nu/J=8.06$ and $\nu/J=5.37$, respectively, as shown in Fig.~\ref{fig:vaet_nonHermitian_spectrum}(a). Notably, the maximum of $P_a(t_f)$ for the two- or three-phonon VAET process appears at a position noticeably shifted away from the EP ($\gamma/J=1.00778\sim1$). 
This shift is ascribed to the change in EP position induced by the vibrational coupling strength $\kappa$, which becomes relatively stronger (as indicated by the ratio $\kappa/\nu= \frac{\kappa/J}{\nu/J}$) for multi-phonon processes with e.g., $2\nu\sim\lambda_{41}=\lambda_{23}$ or $3\nu\sim\lambda_{41}=\lambda_{23}$, for a given dimeric energy structure.  
To further investigate this effect, we have recalculated Fig.~\ref{fig:vaet_nonHermitian_spectrum}(a) with a weaker coupling, specifically with $\kappa/J=0.1$, and observed that there is now no noticeable shift, while the acceptor population is also reduced because of the weaker vibrational coupling. 
Another consequence of this stronger relative coupling strength for the two-phonon VAET process can be observed from the unmarked but clearly visible horizontal line $\nu/J=8.06$ in Fig.~\ref{fig:vaet_nonHermitian_spectrum}(a). This line corresponds to the white dotted line $Jt=22.5$ in Fig.~\ref{fig:vaet_nonHermitian_dynamics}(c) and clearly exhibits more pronounced acceptor population in regions away from the EP than the corresponding population seen for the one-phonon VAET process with $\nu/J=16.12$ [or, equivalently, the line $Jt=22.5$ in Fig.~\ref{fig:vaet_nonHermitian_dynamics}(a)].

Although we have selected $Jt_f=22.5$ as the time point at which $P_a(t)$ reaches its maximum value for $\gamma/J=1$ [see Fig.~\ref{fig:vaet_nonHermitian_dynamics}(b)], we emphasize that the main spectral features of the VAET processes in the presence of the non-Hermiticity that are shown in Fig.~\ref{fig:vaet_nonHermitian_spectrum}(a) are quite universal and independent of the specific values of $t_f$. This universality is demonstrated in Fig.~\ref{fig:vaet_nonHermitian_spectrum}(b), which presents the average population accumulation $\bar{P}_a$ over a time period $t_f =22.5/J$ [see Eq.~(\ref{eq:bar_Pa})]. Like $P_a(t_f)$ in Fig.~\ref{fig:vaet_nonHermitian_spectrum}(a), $\bar{P}_a$ in Fig.~\ref{fig:vaet_nonHermitian_spectrum}(b) also shows two one-phonon VAET processes with almost identical populations in the unbroken symmetry phase to the left of the EP ($\gamma/J \lesssim1$). This similarity is due to the dimer-vibration interaction, which results in optical transitions that are less dependent on $\gamma$, as explained in Sec.~\ref{sec:mechanism} below.

Figures~\ref{fig:vaet_nonHermitian_spectrum}(c) and \ref{fig:vaet_nonHermitian_spectrum}(d) display the phase transition characteristics of the one- and two-phonon VAET processes with $\nu/J=16.12$ and $\nu/J=8.06$, respectively. These plots are constructed from the data in Figs.~\ref{fig:vaet_nonHermitian_spectrum}(a) and \ref{fig:vaet_nonHermitian_spectrum}(b) and constitute an analog of Fig.~\ref{fig:dimer_nonHermitian_dynamics}(d) which represents the corresponding one-dimensional spectra for the dimer without vibration. Importantly, these panels also clearly illustrate the enhancement of VAET near the EP relative to the $\gamma=0$ Hermitian case. 
Regarding the one-phonon VAET process, Fig.~\ref{fig:vaet_nonHermitian_spectrum}(c) demonstrates that the transition from  ${\cal PT}$-symmetry unbroken ($\gamma/J\lesssim1$) to broken ($\gamma/J\gtrsim1$) phases occurs at the EP ($\gamma/J=1.00778\sim1$), 
for both observables, i.e., for $P_a(t_f)$ and the time-averaged $\bar{P}_a$. 
This independence of the specific observable indicates that the VAET spectrum can serve as a reliable means of investigating the ${\cal PT}$-symmetry phase transition, with both $P_a(t_f)$ and $\bar{P}_a$ providing a spectral signature as a function of $\gamma/J$. 
Fig.~\ref{fig:vaet_nonHermitian_spectrum}(d) also shows a much broader lineshape of the two-phonon VAET peak at the quantum phase transition than that seen for the one-phonon VAET peak in Fig.~\ref{fig:vaet_nonHermitian_spectrum}(c).
This difference is also attributed to the coupling strength $\kappa$ being considerably stronger for the two-phonon VAET process ($\nu/J=8.06$, compared to $16.12$ for the one-photon VAET process).

\subsection{Phonon-absorption mechanism \label{sec:mechanism}}

Since in the absence of vibrations we see fast oscillations associated with the excitonic transitions $\lambda_{41}$ and $\lambda_{23}$ [see Fig.~\ref{fig:dimer_nonHermitian_dynamics}(a)], it appears surprising that in the presence of vibrations neither Fig.~\ref{fig:vaet_nonHermitian_spectrum}(a) nor Fig.~\ref{fig:vaet_nonHermitian_spectrum}(b) shows any dominant feature of a corresponding one-phonon VAET process. In particular, there is no evidence of any spectral feature corresponding to enhancement of $P_a(t_f)$ or $\bar{P}_a$ when the vibrational frequency is equal to the excitonic transition energy, i.e., $\nu=\lambda_{41}=\lambda_{23}=16.12J$ [represented by the cyan up-down arrows in Fig.~\ref{fig:schematic}(a)]. This absence is most clearly evident both in the Hermitian limit ($\gamma=0$) and in the unbroken symmetry phase away from the EP.  Nevertheless, there is a strong spectral feature for this vibrational frequency at the location of the EP, which we shall discuss below, and also a very weak feature in the broken symmetry phase. 

The absence of a clear one-phonon feature away from the EP can be attributed to the donor-vibration interaction represented by $H_{\rm int}\equiv\kappa\sigma_z^{(d)}(a+a^{\dagger})$ in Eq.~(\ref{eq:H}), which leads to nearly vanishing transition matrix elements between eigenstates $|\psi_{4(2)}\rangle$ and $|\psi_{1(3)}\rangle$. % 
Given that the vibration is initially in the thermal state (e.g., $k_BT/J=40$), this behavior can be understood analytically by examining the transition matrix elements of $H_{\rm int}$ between the eigenstates of the decoupled donor and acceptor in the Hermitian limit, namely $|\psi_4\rangle \rightarrow \frac{|ge\rangle+|ee\rangle}{\sqrt{2}}$ and $|\psi_1\rangle\rightarrow\frac{|gg\rangle+|eg\rangle}{\sqrt{2}}$ or $|\psi_2\rangle \rightarrow \frac{|ge\rangle -|ee\rangle}{\sqrt{2}}$ and $|\psi_3\rangle\rightarrow \frac{|gg\rangle -|eg\rangle}{\sqrt{2}}$ [see Eqs.~(\ref{eq:psi_13_limit}) and (\ref{eq:psi_42_limit}) as well as the grey dashed lines in Fig.~\ref{fig:schematic}(a)].

\begin{figure}%[hbt!]
\centering
  \includegraphics[width=.86\columnwidth]{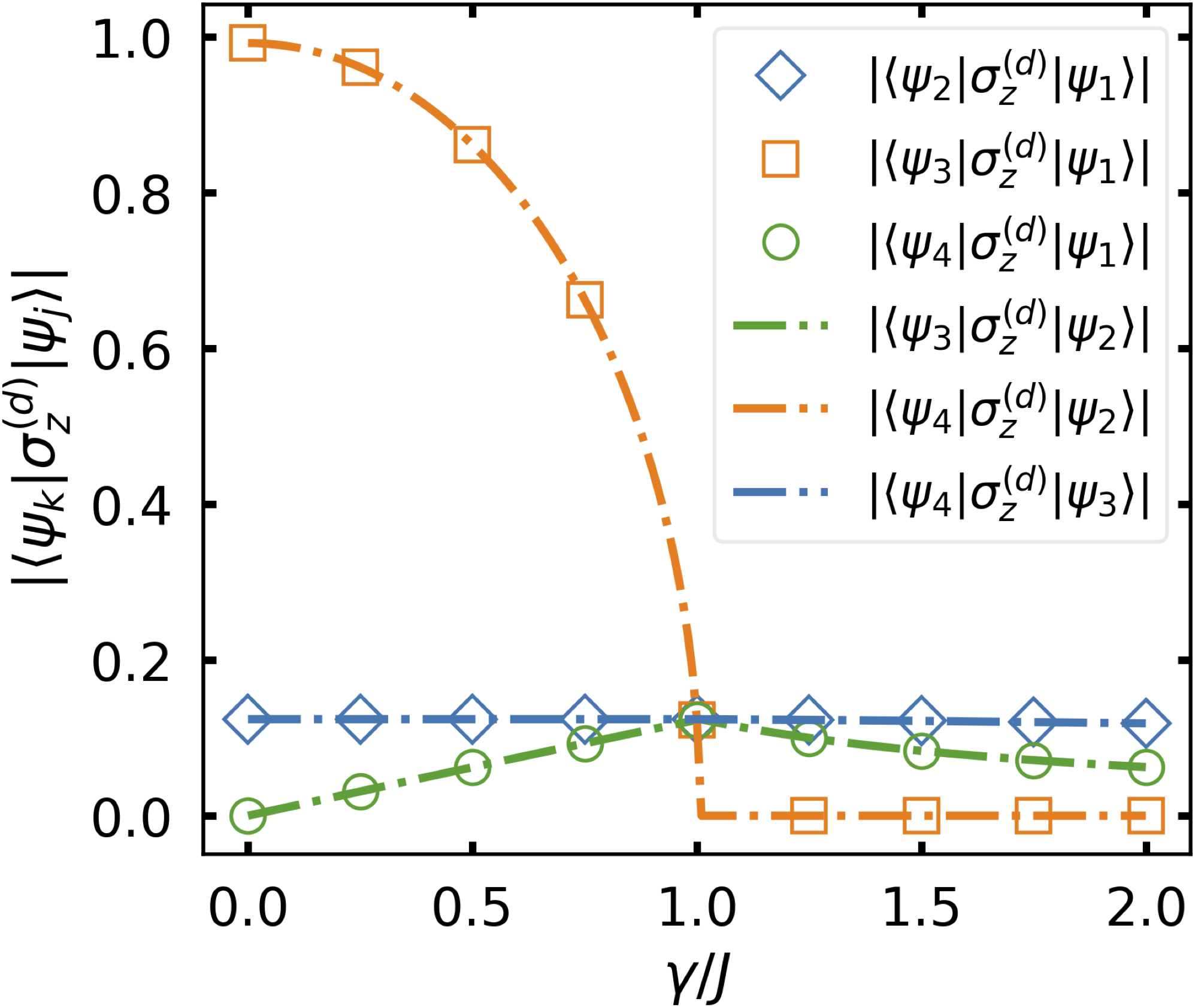}
\caption{(color online) Transition matrix elements of the donor operator $\sigma_z^{(d)}$ in the donor-vibration interaction [i.e., $H_{\rm int}\equiv\kappa\sigma_z^{(d)}(a+a^{\dagger})$ in Eq.~(\ref{eq:H})] between eigenstates $|\psi_j\rangle$ and $|\psi_{k(\neq j)}\rangle$ ($j,k=1,2,3,4$) of the non-Hermitian dimer given in Eq.~(\ref{eq:H_dimer}). 
The parameters used are $\alpha/J=1$ and $\Delta/J=8$ and the corresponding two-fold degenerate second-order EP is at $\gamma/J=1.00778\sim1$.}
\label{fig:ABSmtrelePsi_ij}
\end{figure}

To gain more physical understanding, we numerically calculate the transition matrix elements of the donor operator $\sigma_z^{(d)}$ in $H_{\rm int}$ between the eigenstates $|\psi_j\rangle$ and $|\psi_{k\neq j}\rangle$ of Eq.~(\ref{eq:H_dimer}). These are presented in Fig.~\ref{fig:ABSmtrelePsi_ij}. 
The green circles with 
the dashed-dotted curve show that the transitions between $|\psi_1\rangle$ and $|\psi_4\rangle$ and between $|\psi_2\rangle$ and $|\psi_3\rangle$, respectively, are forbidden in the Hermitian limit, implying that there is no VAET at $\gamma=0$, 
while these transitions become maximally favorable at the EP. This explains why the one-phonon VAET process with $\nu/J=16.12$ in Figs.~\ref{fig:vaet_nonHermitian_spectrum}(a) and \ref{fig:vaet_nonHermitian_spectrum}(b) is not prominent in the region away from the EP. 
At the same time, the orange squares with the dashed-dotted curve indicate that in the symmetry unbroken phase, the vibration induces instead transitions between $|\psi_1\rangle$ and $|\psi_3\rangle$, and between $|\psi_2\rangle$ and $|\psi_4\rangle$ with $\lambda_{13(42)}=\lambda_{1(4)}-\lambda_{3(2)}$. This frequency is approximately $1.217J$ for $\gamma/J=0.8$, which corresponds to the frequency of oscillations for the orange curve in Fig.~\ref{fig:dimer_nonHermitian_dynamics}(a) that are induced in the Hermitian system by addition of the non-Hermitian gain/loss $\gamma$. However, this small value of vibrational frequency $\nu$ implies a relatively large average phonon number that is well beyond region of validity of the calculations in Figs.~\ref{fig:vaet_nonHermitian_spectrum}(a) and \ref{fig:vaet_nonHermitian_spectrum}(b) (see Sec.~\ref{sec:model}). 
Finally, the blue diamonds with the dashed-dotted curve show the corresponding matrix elements for vibrationally induced transitions between $|\psi_1\rangle$ and $|\psi_2\rangle$, and between $|\psi_3\rangle$ and $|\psi_4\rangle$. These correspond to the one-phonon VAET processes starting from $\nu/J=14$ and $\nu/J=18$, respectively, that are observed in the top left quadrants of Figs.~\ref{fig:vaet_nonHermitian_spectrum}(a) and \ref{fig:vaet_nonHermitian_spectrum}(b). 
The lack of any dependence of $|\langle\psi_2|\sigma_z^{(d)}|\psi_1\rangle|$ (blue diamonds) and $|\langle\psi_4|\sigma_z^{(d)}|\psi_3\rangle|$ (blue dashed-dotted curve) on $\gamma$ accounts for the fact that these two one-phonon VAET processes have nearly identical intensities [see, e.g, the top left quadrant of Fig.~\ref{fig:vaet_nonHermitian_spectrum}(b)].

\begin{figure}%[hbt!]
\centering
  \includegraphics[width=.83\columnwidth]{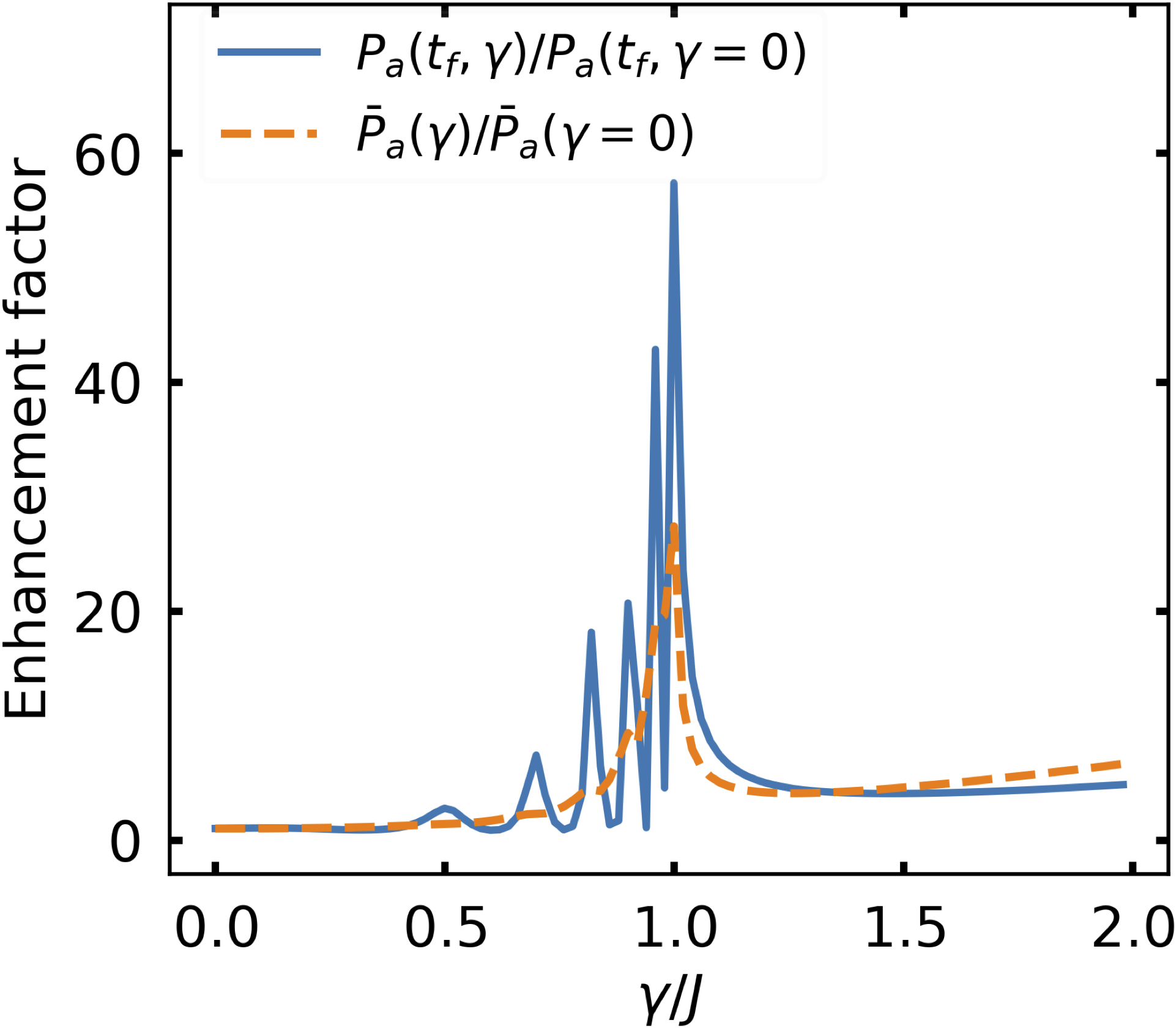}
\caption{(color online) Enhancement factor for VAET of the non-Hermitian system, defined as the ratio of the population $P_a(t_f,\gamma)$ or $\bar{P}_a(\gamma)$ to the corresponding Hermitian population $P_a(t_f,\gamma=0)$ or $\bar{P}_a(\gamma=0)$, respectively. 
The two-fold degenerate second-order EP is located at $\gamma/J=1.00778\sim1$. The parameters used here are $\alpha/J=1$, $\Delta/J=8$, $t_f=22.5/J$, $\kappa/J=0.3$, $k_BT/J=40$, and the initial state is $|eg\rangle$.
}
\label{fig:enhancementFactor}
\end{figure}

Figure~\ref{fig:ABSmtrelePsi_ij} shows that at $\gamma/J\sim1$ the four transition matrix elements
$|\langle\psi_2|\sigma_z^{(d)}|\psi_1\rangle|$,  $|\langle\psi_4|\sigma_z^{(d)}|\psi_3\rangle|$, 
$|\langle\psi_4|\sigma_z^{(d)}|\psi_1\rangle|$, and  $|\langle\psi_3|\sigma_z^{(d)}|\psi_2\rangle|$ 
%at $\gamma/J\sim1$ 
are identical in magnitude, resulting in four simultaneous transitions (with $\lambda_{41}=\lambda_{23}=\lambda_{21}=\lambda_{43}$ for a given dimeric energy structure with the parameters $\alpha/J=1$ and $\Delta/J=8$) that can be resonantly excited by a single phonon from the vibrational mode with $\nu/J=16.12$. 
This results in the maximal population, i.e., $P_a(t_f)=0.895$ and $\bar{P}_a=0.248$, seen at the EP for the one-phonon VAET processes in Figs.~\ref{fig:vaet_nonHermitian_spectrum}(a) and \ref{fig:vaet_nonHermitian_spectrum}(b). 
To quantify the enhancement of vibrationally assisted energy transfer relative to the Hermitian case, we analyze the population $P_a(t_f)$ or $\bar{P}_a$ of the non-Hermitian one-phonon VAET process relative to its corresponding Hermitian case. The enhancement factor can be defined as $P_a(t_f,\gamma)/P_a(t_f,\gamma=0)$ or $\bar{P}_a(\gamma)/\bar{P}_a(\gamma=0)$ and is plotted in
Fig.~\ref{fig:enhancementFactor}. This shows that as the EP at $\gamma/J=1.00778\sim1$ is approached, the non-Hermitian one-phonon VAET process is significantly enhanced compared to the Hermitian case, by a factor up to approximately $57$ for $P_a(t_f,\gamma)/P_a(t_f,\gamma=0)$ (blue solid curve), and up to $27$ for $\bar{P}_a(\gamma)/\bar{P}_a(\gamma=0)$ (orange dashed curve).

This analysis of the mechanism for phonon absorption highlights the simultaneous excitation of multiple one-phonon VAET processes at an EP due to the unique non-Hermitian degeneracy in this case. 
It is important to note that the presence of the two-fold degenerate second-order EP is crucial, since it is this that allows for the occurrence of four simultaneous excitations. The phenomenon arises from the fact that these simultaneous transitions occur between eigenstates associated with  distinct EPs.  
In general, if we consider an $s$-fold degenerate $n$th-order EP, we would have a total of $C_1^sC_1^nC^{s-1}_1C^n_1/2=s(s-1)n^2/2$ simultaneous transitions between two  
eigenstates associated with distinct EPs, where ${\cal N} \geq s, n >1$  with ${\cal N}$ the total number of eigenstates and $C^{p}_{q}=\frac{p!}{q!(p-q)!}$ are binomial coefficients. For the current work with $s=n=2$ and ${\cal N}=4$, this reduces to 4. 
When there is no EP degeneracy at the $n$th-order EP, for example, if $n={\cal N}$, as in the case of a fourth-order EP in a two-qubit non-Hermitian system~\cite{LiChenWhaley22prl}, there would be no simultaneous transitions between 
%non-coalesced 
eigenstates associated with distinct EPs, resulting in no absorption of phonons. However, if there is no degeneracy and the order of the EP is less than the number of eigenstates, i.e., $1<n<{\cal N}$, each of the $n$ coalesced eigenstates can undergo transitions to the ${\cal N}-n$ additional eigenstates, resulting in $n$ simultaneous excitations resonantly induced by a single phonon, equal to the $n$th-order nature of the EP. 
This implies that for a non-degenerate EP, a larger system than a dimer would be required in order to achieve the four simultaneous excitations necessary for the VAET enhancement that is observed in the current work.  
The significance of the degeneracy of the EP is also evident when considering that a non-degenerate second-order EP, which allows only two simultaneous excitations, is anticipated to result in a weaker enhancement compared to what we see here in the case of the two-fold degenerate second-order EP with four simultaneous excitations. 
The resulting excitation process is fundamentally different from not only the corresponding excitation processes in the presence of Hermitian degeneracy, but also from the coherent excitation of two atoms of identical frequency by one photon~\cite{GarzianoNoriSavasta16prl}.

\begin{figure}%[hbt!]
\centering
  \includegraphics[width=0.99\columnwidth]{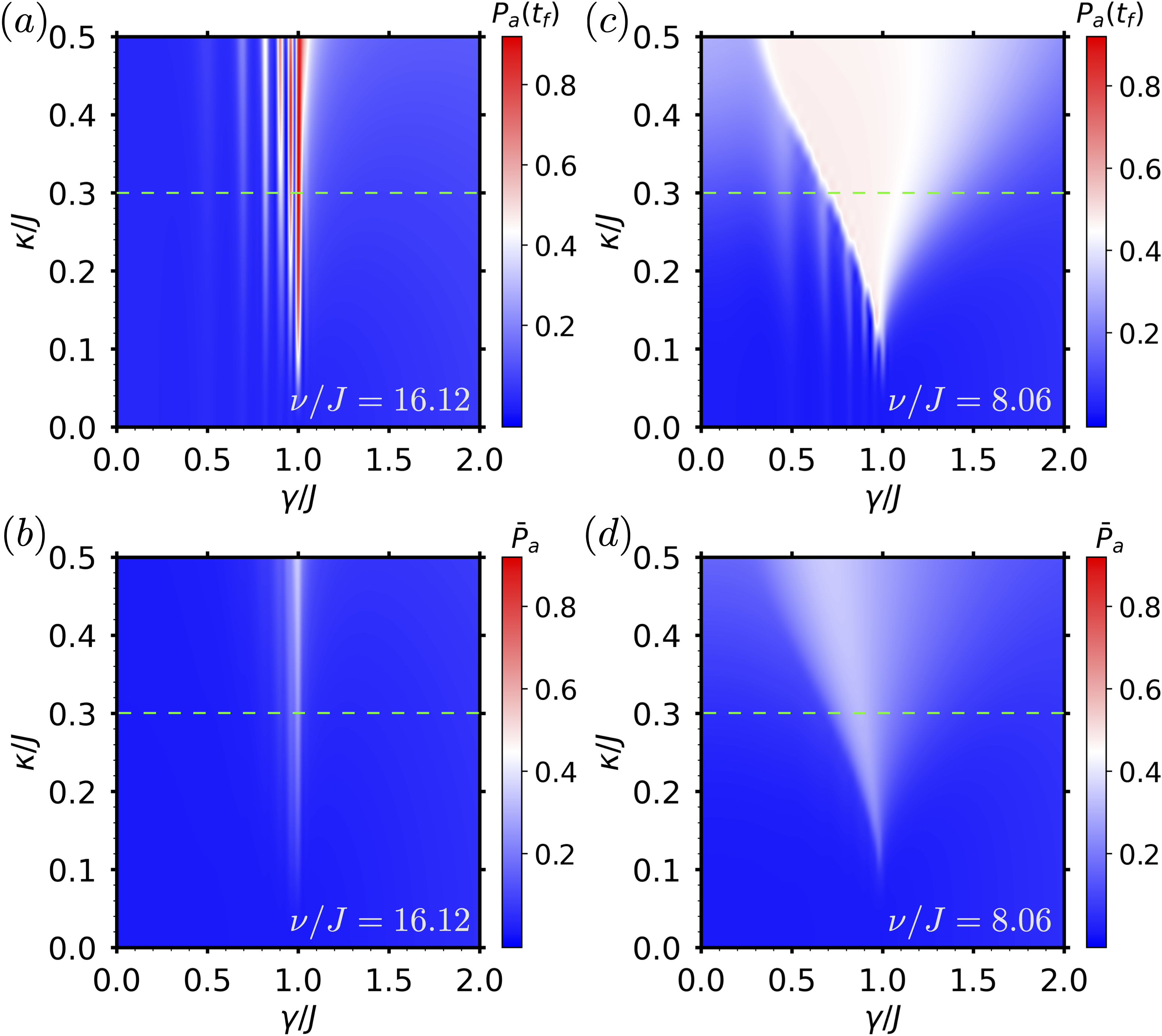}
\caption{(color online) Dependence of the non-Hermitian VAET features quantified by $P_a(t_f)$ [panels (a) and (c)] and by $\bar{P}_a$ [panels (b) and (d)] on the exciton-phonon coupling strength $\kappa$. One-phonon processes with $\nu/J=16.12$ are presented in panels (a) and (b). Two-phonon processes with $\nu/J=8.06$ are presented in panels (c) and (d). The horizontal dashed line $\kappa/J=0.3$ in panels (a) and (c) correspond to the location of the spectral features at $\nu/J=16.12$ and $\nu/J=8.06$ in Fig.~\ref{fig:vaet_nonHermitian_spectrum}(a), respectively, while the dashed lines in panels (b) and (d) correspond to the spectral features in Fig.~\ref{fig:vaet_nonHermitian_spectrum}(b). We use the parameters $\alpha/J=1$ and $\Delta/J=8$, for which the two-fold degenerate second-order EP is located at $\gamma/J=1.00778\sim1$. Other parameters used are $t_f=22.5/J$ and $k_BT/J=40$. The initial state is $|eg\rangle$ in all calculations.}
\label{fig:Ptf_Pint_vs_gamma_kappa}
\end{figure}

\section{Robustness \label{sec:robustness}}

The enhancement of VAET processes in a non-Hermitian quantum system discussed above is based on two considerations: a fixed dimer-vibration coupling strength ($\kappa/J=0.3$) and a vibrational temperature ($k_BT/J=40$) that is sufficiently high for the vibrational mode to provide the required one or two phonons [$1/(e^{\nu/k_BT}-1)\gtrsim 2$] to facilitate energy transfer in the chromophore dimer. 
In order to investigate the robustness of the observed enhancement with regard to these assumptions, we analyze the response of the one- and two-phonon VAET signals to variations in the coupling strength $\kappa$ and temperature $k_BT$.

Figure~\ref{fig:Ptf_Pint_vs_gamma_kappa} shows that
on gradually increasing $\kappa/J$ from zero to $0.5$ while passing through the value of $0.3$ used in the previous section, both the acceptor population [$P_a(t_f=22.5/J)$ in Fig.~\ref{fig:Ptf_Pint_vs_gamma_kappa}(a)] and the corresponding average population accumulation during this time period [$\bar{P}_a$ in Fig.~\ref{fig:Ptf_Pint_vs_gamma_kappa}(b)] exhibit an enhancement of the one-phonon VAET process ($\nu/J=16.12$) near the EP (located at $\gamma/J\sim1$).  
We note that the oscillations in $\gamma$ observed for the one-phonon processes in panel (a) are more pronounced than those for the two-phonon processes in panel (b). 
For a given value of coupling strength $\kappa$, such as the green dashed line at $\kappa/J=0.3$, $P_a(t_f)$ exhibits oscillations with increasing peak amplitude on approaching the EP, consistent with the observations in Figs.~\ref{fig:vaet_nonHermitian_spectrum}(a) and \ref{fig:vaet_nonHermitian_spectrum}(b). 
In comparison, the two-phonon VAET process with $\nu/J=8.06$ in Fig.~\ref{fig:Ptf_Pint_vs_gamma_kappa}(c) [$P_a(t_f)$] or Fig.~\ref{fig:Ptf_Pint_vs_gamma_kappa}(d) [$\bar{P}_a$] shows a smaller enhancement over a broad range of $\gamma/J$ near the EP at $\gamma/J=1.00778\sim1$. This different response of the two-phonon VAET process to varying $\kappa$ is consistent with the behavior seen in Figs.~\ref{fig:vaet_nonHermitian_dynamics} and \ref{fig:vaet_nonHermitian_spectrum}.

\begin{figure}%[hbt!]
\centering
  \includegraphics[width=0.98\columnwidth]{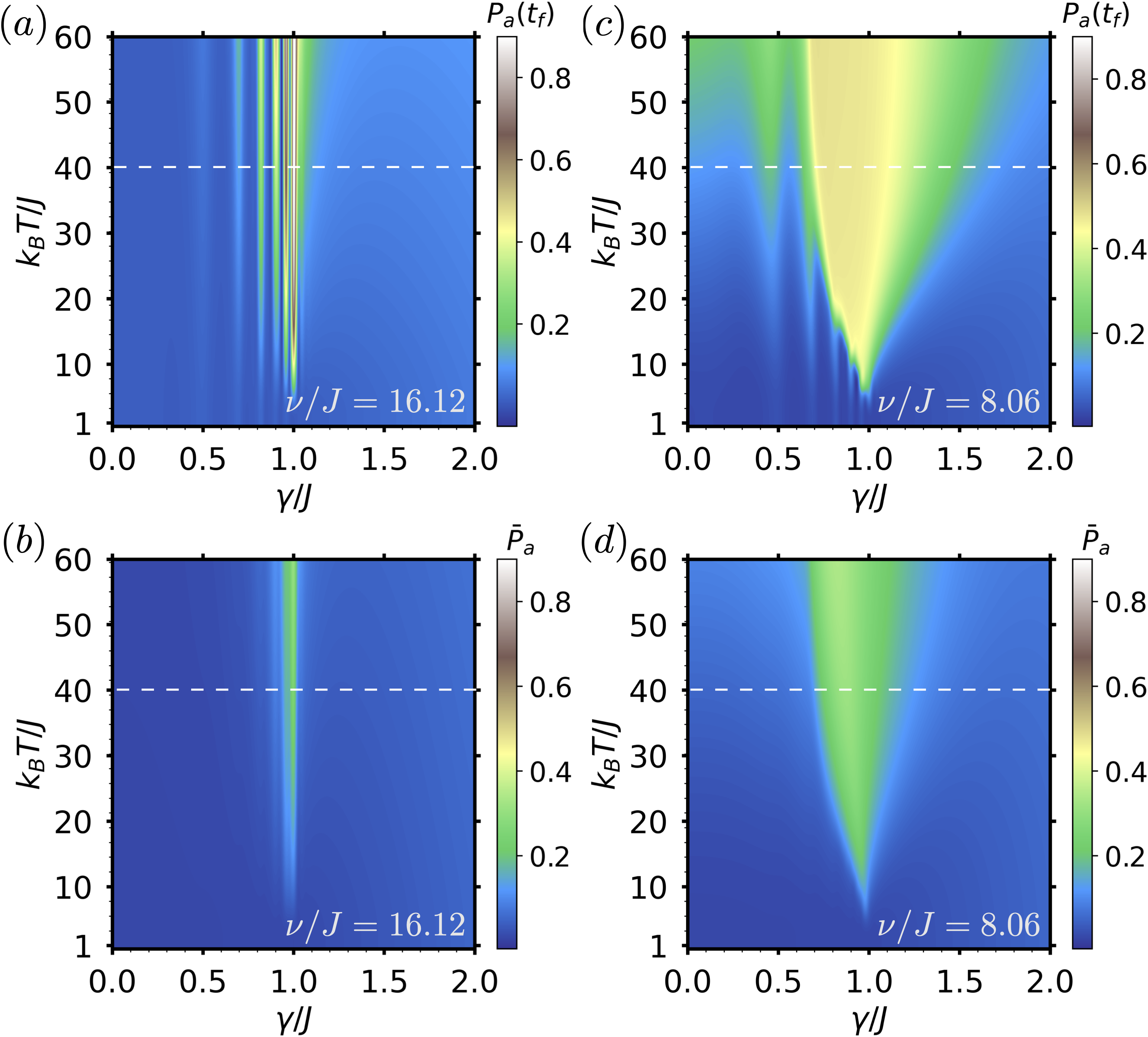}
\caption{(color online) Dependence of the non-Hermitian VAET features quantified by $P_{a}(t_f)$ [panels (a) and (c)] and $\bar{P}_a$ [panels (b) and (d)] on the temperature $k_BT$ of the vibration. One-phonon processes with $\nu/J=16.12$ are presented in panels (a) and (b). Two-phonon processes with $\nu/J=8.06$ are presented in panels (c) and (d). The horizontal dashed line $k_BT/J=40$ in panels (a) and (c) corresponds to the location of the spectral features at $\nu/J=16.12$ and $\nu/J=8.06$ in Fig.~\ref{fig:vaet_nonHermitian_spectrum}(a), respectively, while the dashed line in panels (b) and (d) correspond to the spectral features in Fig.~\ref{fig:vaet_nonHermitian_spectrum}(b). We use the parameters $\alpha/J=1$ and $\Delta/J=8$, for which the two-fold degenerate second-order EP is located at $\gamma/J=1.00778\sim1$. Other parameters used are $t_f=22.5/J$ and $\kappa/J=0.3$. The initial state is $|eg\rangle$ in all calculations.} 
\label{fig:Ptf_Pint_vs_gamma_kT}
\end{figure}

Figure~\ref{fig:Ptf_Pint_vs_gamma_kT} summarizes the response of the acceptor population $P_a(t_f)$ and 
the average population accumulation $\bar{P}_a$ 
when varying the temperature $k_BT$, while keeping the coupling strength fixed at $\kappa/J=0.3$. 
It is evident that the enhancement of the one-phonon VAET process ($\nu/J=16.12$) near the EP for both $P_a(t_f)$ [panel (a)] and $\bar{P}_a$ [panel (b)] can be further amplified by increasing the temperature. 
In the case of the two-phonon VAET processes ($\nu/J=8.06$), both panel (c) for $P_a(t_f)$, and panel (d) for $\bar{P}_a$, also clearly demonstrate an enhancement over a broad range of $\gamma/J$ values around the EP, but with a significantly smaller population in the acceptor than that achieved by the one-phonon VAET. 
However, while the enhanced VAET processes in Fig.~\ref{fig:Ptf_Pint_vs_gamma_kT} appear qualitatively similar to those in Fig.~\ref{fig:Ptf_Pint_vs_gamma_kappa}, the underlying mechanisms of the enhancement are different. In Fig.~\ref{fig:Ptf_Pint_vs_gamma_kT} the increased enhancement derives from the availability of a greater number of phonons available at high temperatures to facilitate the transfer process, while in Fig.~\ref{fig:Ptf_Pint_vs_gamma_kappa} the enhancement derives from the stronger exciton-phonon coupling.

\begin{figure}%[hbt!]
\centering
  \includegraphics[width=.99\columnwidth]{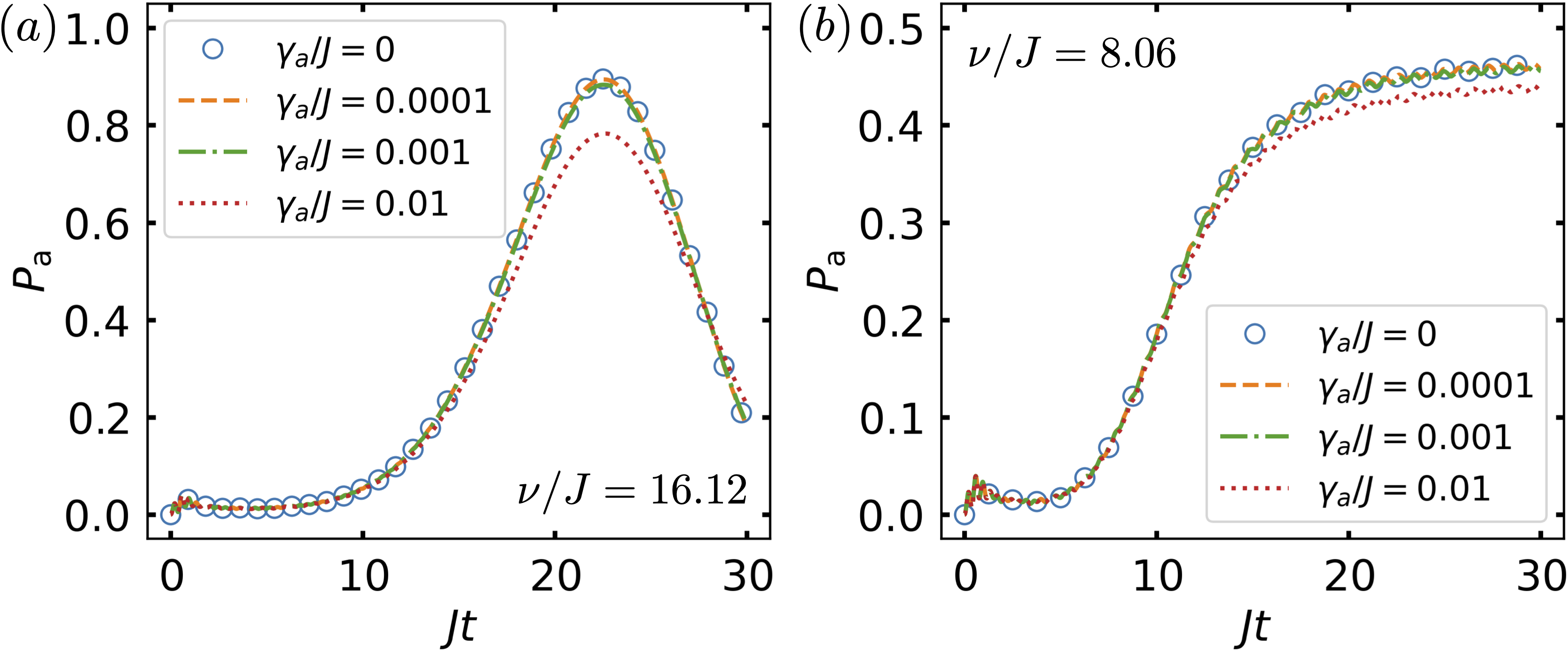}
\caption{(color online) The acceptor population for the one- and two-phonon VAET processes presented in (a) and (b), respectively, calculated via the master equation in Eq.~(\ref{eq:ME}) for several values of dissipation rate: $\gamma_a/J=0,\, 0.0001,\, 0.001,\, 0.01$ where $\gamma_a=0$ (blue circles, same as $\gamma/J=1$ in Fig.~\ref{fig:vaet_nonHermitian_dynamics}) is included for reference. Other parameters used are $\alpha/J=1$, $\Delta/J=8$, $k_BT/J=40$, $\kappa/J=0.3$, and $\gamma/J=1$. The initial state is $|eg\rangle$ in all calculations.}
\label{fig:veat_nonHermitian_lindblad}
\end{figure}

In addition to the loss included in the non-Hermitian Hamiltonian, we have also considered the decohering effect of spontaneous emission at the acceptor. To incorporate this effect, we numerically solve the Lindblad master equation %given by
\begin{eqnarray}
\frac{d\varrho}{dt} &=& -i\left(H\varrho-\varrho H^{\dagger} \right) 
+\gamma_a \left(\sigma_-^{(a)}\varrho\sigma_+^{(a)} 
 -\frac{1}{2}\left\{\sigma_+^{(a)}\sigma_-^{(a)}, \varrho \right\}\right) ,  \notag\\
\label{eq:ME}
\end{eqnarray}
with $H$ defined in Eq.~(\ref{eq:H}), and the Lindblad operator $\sigma_-^{(a)}=|g\rangle_a\langle e|$. The resulting total density matrix $\rho=\varrho/{\rm Tr}[\varrho]$ reduces to Eq.~(\ref{eq:rho_total}) when $\gamma_a=0$. 
Figure~\ref{fig:veat_nonHermitian_lindblad} shows the time evolution of acceptor population $P_a(t)$ [Eq.~(\ref{eq:Pa_t})] obtained from Eq.~(\ref{eq:ME}) for the one- and two-phonon VAET processes presented in panel (a) and panel (b), respectively.
Comparing with the corresponding non-Hermitian dynamics for $\gamma_a=0$ (the circles in Fig.~\ref{fig:veat_nonHermitian_lindblad}), it is evident that $P_a(t)$ is not severely suppressed by spontaneous emission from the acceptor. This implies that these EP-enhanced VAET processes in the non-Hermitian quantum system should be observable under a reasonable amount of dissipation, as long as $\gamma_a\ll J,\, \Delta$.

\section{Summary and Conclusions \label{sec:conclusion}}

We have investigated vibrationally assisted energy transfer processes in a non-Hermitian quantum system involving a ${\cal PT}$-symmetric chromophore dimer weakly coupled to a vibrational mode. We first demonstrated the existence of EPs and the non-Hermitian features, such as slow oscillations induced by gain and loss in the ${\cal PT}$-symmetry unbroken phase and non-equilibrium steady states in the broken phase, for the excitation energy transfer processes in the absence of coupling to vibrational modes. 
These EPs are two-fold degenerate second-order EPs, which has significant implications for both the non-Hermitian dynamics and the dimer spectra. 
Adding the vibrational mode, we found that both one- and two-phonon VAET processes were enhanced by the presence of the non-Hermiticity. The dynamical and spectral features of these enhanced VAET processes were then analyzed. 
The enhancement near the EP could be attributed to the maximally favorable phonon absorption occurring at the two-fold degenerate second-order non-Hermitian degeneracy point, where four simultaneous transitions between 
%non-coalesced 
eigenstates associated with distinct EPs are excited by a single phonon. This generates a unique spectral feature that indicates the coalescing of multiple eigenstates in addition to the degeneracy of the eigenenergies, providing a new way to probe exceptional points and ${\cal PT}$-symmetry phase transitions. The results were found to be robust to variations in the exciton-phonon coupling, to the phonon temperature, and to the presence of spontaneous emission on the acceptor.

Our proposed novel approach of probing EPs via fluorescence-detected vibrational spectroscopy, using the acceptor population measured through fluorescence detection in trapped-ion experiments in place of the traditional spectral amplitude at a given vibrational frequency, offers several advantages. While both approaches can indicate the amount of phonon emission at that specific frequency,  
the acceptor population method provides valuable insights into the coalescence of eigenstates or eigenenergies at the EP. 
By continuously scanning the frequency of the vibration, we have shown that one can effectively map and observe the process of eigenstates or eigenenergies converging at the EP. 
Additionally, by comparing the population data with the spectral amplitude observed in the traditional spectral experiments, one can establish crucial correlations, 
shedding light on the population transfer processes occurring at the EP. 
In particular, rapid increases or decreases in the acceptor population, as monitored by fluorescence measurements, imply that there is a corresponding peak or
dip in the spectral amplitude data at the precise vibrational frequency that would be observed by traditional spectroscopy experiments, which are however challenging to carry out in this setting. 
In summary, the advantages of utilizing a fluorescence-detected vibrational spectroscopy to probe exceptional points include its capability to map the coalescence of eigenstates or eigenenergies, its sensitivity to non-Hermitian effects, and its capacity to offer significant insights into the dynamics and behavior of the system in close proximity to the EP. 

In contrast to Hermitian degeneracy, where degenerate eigenstates are always linearly independent, it is well-known that at exceptional points in a non-Hermitian system, the eigenstates coalesce and become degenerate in a nontrivial manner, as is shown for our dimer system in Fig.~\ref{fig:dimer_nonHermitian}(d). 
For a ${\cal PT}$-symmetric system, this nontrivial coalescence is associated with breaking of the ${\cal PT}$ symmetry at the exceptional point of the non-Hermitian system. 
Furthermore, the eigenstates are also continuously dependent on the non-Hermitian parameter combination of $J$ and $\gamma$. This results in complex and unique non-Hermitian dynamics within both the ${\cal PT}$-symmetric unbroken ($0 < \gamma/J < 1$) and broken ($\gamma/J >1$) phases,
as depicted in Fig.~\ref{fig:dimer_nonHermitian_dynamics}(a, b, d) and in Fig.~\ref{fig:vaet_nonHermitian_dynamics}(a-d) in the presence and absence of the vibrational mode, respectively.
A remarkable feature of the dimer system studied in this work is that the two pairs of coalesced eigenstates (i.e., $|\psi_1\rangle$ with $|\psi_3\rangle$, and $|\psi_2\rangle$ with $|\psi_4\rangle$) result in four transition matrix elements that are identical in magnitude at the EP, as shown explicitly in Fig.~\ref{fig:ABSmtrelePsi_ij}.
The values of these matrix elements play a crucial role in the acceptor population for the one-phonon VAET process. This degeneracy of the matrix elements at the EP accounts for the observation that the one-phonon VAET process with $\nu/J=16.12$ is maximized at the EP,  
with the acceptor population $P_a(t_f)$ showing a marked increase over the corresponding values in both the unbroken and broken ${\cal PT}$ symmetry phases (see
Fig.~\ref{fig:vaet_nonHermitian_spectrum}). This behavior deriving from the coalescence of the eigenvectors is not seen in the $\gamma=0$ Hermitian limit.

Realization of the non-Hermitian VAET phenomena reported here is expected to be possible for trapped ions in the near term, given recent advancements in both Hermitian VAET experiments with trapped ions~\cite{GormanHaeffner18prx} and the realization of non-Hermitian trapped-ion qubits~\cite{DingZhang21prl,WangJingChen21pra,quinn2023observing}. 
Implementing the non-Hermitian VAET involves encoding the excitonic states in individual ions and adjusting the Hamiltonian parameters of tunneling coupling, excitonic donor-acceptor interaction, and donor-vibration coupling, by coherent laser drives, as well as manipulation of the non-Hermitian gain and loss terms by the heating and cooling techniques outlined in Sec.~\ref{sec:model}. The vibrational frequencies can be controlled and tuned via electromagnetic fields, as described in Sec.~\ref{sec:model}.

In addition to the trapped-ion platform, it is of interest to explore other physical systems for experimental realization of these findings of enhanced VAET processes near an EP. For instance, in the case of superconducting circuits, non-Hermitian systems can be achieved by using systems of qutrits with post-selection~\cite{NaghilooMurch19nphys}.

\acknowledgements 
We thank Mohan Sarovar and Liwen Ko for helpful discussions in the early stages of this work. Z.Z.L also thanks Weijian Chen for insightful comments and suggestions. 
This work has been supported by AFOSR MURI Grant No. FA9550-21-1-0202 and by the U.S. Department of Energy, Office of Science, Office of Basic Energy Sciences under Award Number DE-SC0023277.

\appendix

\section{${\cal PT}$-symmetric non-Hermitian donor \label{app:donor}}

The non-Hermitian donor in the chromophore dimer is modeled as a two-level system subject to gain and loss, and its Hamiltonian is described by   
\begin{eqnarray}
H_{d} &=& -i\gamma\sigma_z^{(d)} + J\sigma_x^{(d)},
\label{eq:H_donor}
\end{eqnarray}
where $\sigma_x^{(d)}=|g\rangle_d\langle e| + |e\rangle_d\langle g|$ and $\sigma_z^{(d)}=|e\rangle_d\langle e| - |g\rangle_d\langle g|$. It is easy to verify that the non-Hermitian Hamiltonian in Eq.~(\ref{eq:H_donor}) respects \sout{the} ${\cal PT}$-symmetry, i.e., ${\cal PT}H_{d}{\cal PT} = H_{d}$ with ${\cal P}=\sigma_x^{(d)}$ and ${\cal T}$ being a complex conjugation. 
The corresponding eigenenergies and eigenstates are obtained as $\lambda_{1,2}=\mp\sqrt{J^2-\gamma^2}$ and
\begin{equation}
|\tilde{\varphi}_1\rangle =  \begin{pmatrix}
   -\frac{\sqrt{J^2-\gamma^2}+i\gamma}{J} \\
   1 
   \end{pmatrix}, 
|\tilde{\varphi}_2\rangle =  \begin{pmatrix}
   \frac{\sqrt{J^2-\gamma^2}-i\gamma}{J} \\
   1 
   \end{pmatrix} ,
\label{eq:donorEigenstates}
\end{equation} 
respectively. 
The normalized eigenstates are then formally given by $|\varphi_1\rangle=\frac{|\tilde{\varphi}_1\rangle}{||\tilde{\varphi}_1\rangle|}=(\varphi_{1,0},\varphi_{1,1})^T$ and $|\varphi_2\rangle=\frac{|\tilde{\varphi}_2\rangle}{||\tilde{\varphi}_2\rangle|}=(\varphi_{2,0},\varphi_{2,1})^T$.

\begin{figure}%[hbt!]
\centering
  \includegraphics[width=.96\columnwidth]{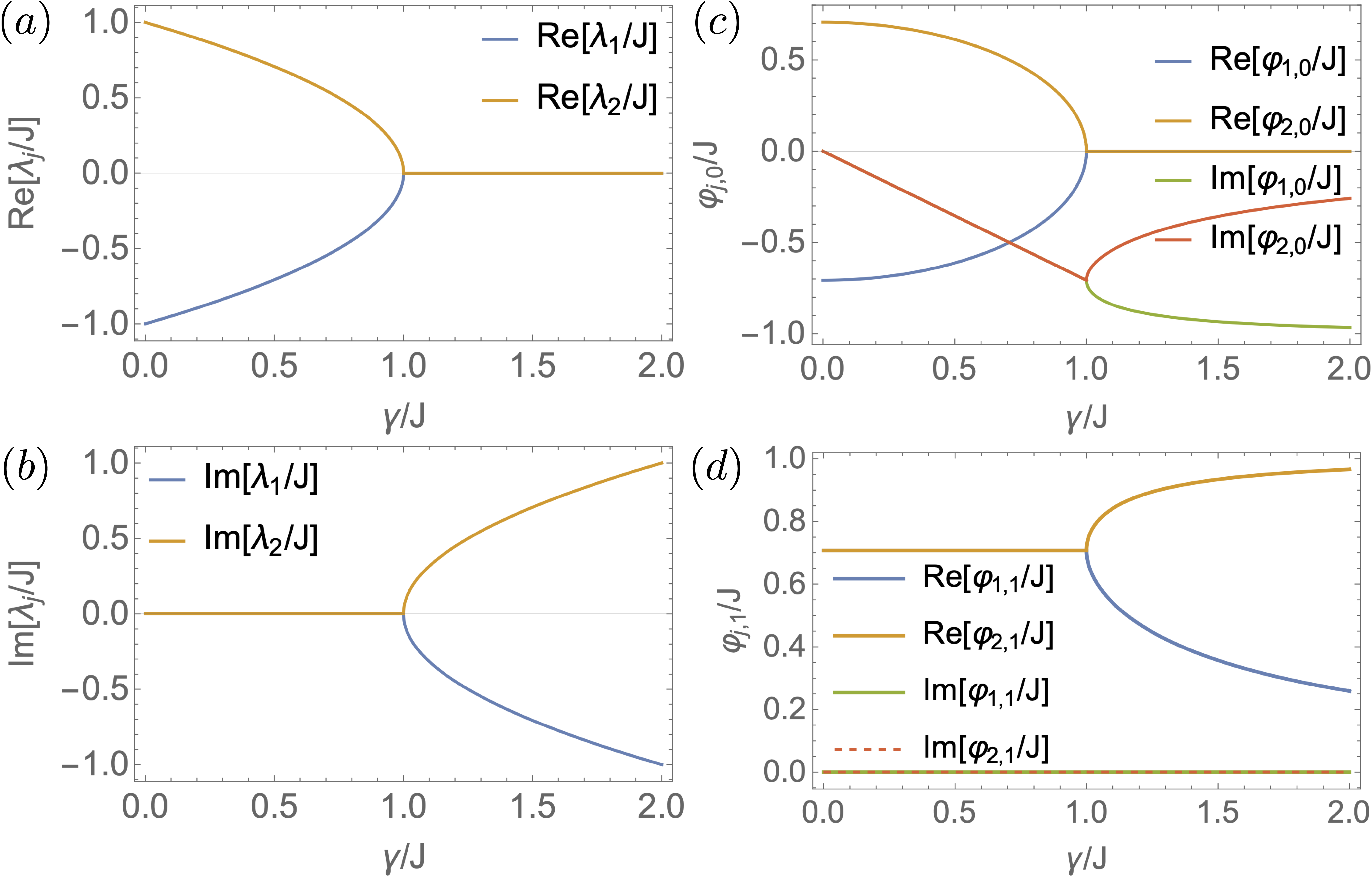}
\caption{(color online) Left panel: Real and imaginary parts of the eigenenergies $\lambda_j$ presented in (a) and (b) respectively. Right panel: The first elements $\varphi_{j,0}$ and second elements $\varphi_{j,1}$ of the eigenstates $|\varphi_j\rangle=(\varphi_{j,0},\varphi_{j,1})^T$ ($j=1,\,2$) presented in (c) and (d) respectively.}
\label{fig:donor_eigen}
\end{figure}

The real and imaginary parts of each eigenenergy are presented in Figs.~\ref{fig:donor_eigen}(a) and \ref{fig:donor_eigen}(b), respectively. It is shown that eigenenergies $\lambda_{1,2}$ in Figs.~\ref{fig:donor_eigen}(a) and \ref{fig:donor_eigen}(b) are both real when $\gamma/J<1$ in the unbroken symmetry phase, indicating both levels of the donor are coherently populated. These two eigenenergies further coalesce at $\gamma/J=1$, signifying a second-order EP. When $\gamma/J>1$, i.e., in the symmetry broken phase, the eigenenergies become purely imaginary, and correspondingly the population at $|e\rangle_d$ decreases exponentially while the population at the level $|g\rangle_g$ increases. 
Figure~\ref{fig:donor_eigen}(c) shows the first elements, i.e., $\varphi_{1,0}$ and $\varphi_{2,0}$, of normalized eigenstates $|\varphi_{1}\rangle$ and $|\varphi_{2}\rangle$, respectively, with real and imaginary parts presented separately. Figure~\ref{fig:donor_eigen}(d) demonstrates the second elements, i.e., $\varphi_{1,1}$ and $\varphi_{2,1}$, of the normalized eigenstates. It is shown that the eigenstates of the non-Hermitian donor coalesce at the EP of the second order, i.e., when $\gamma/J=1$.

\section{Relation between donor-acceptor energy levels and conditions for uphill energy transfer \label{sec:uphill}}

\begin{figure}%[hbt!]
\centering
  \includegraphics[width=.96\columnwidth]{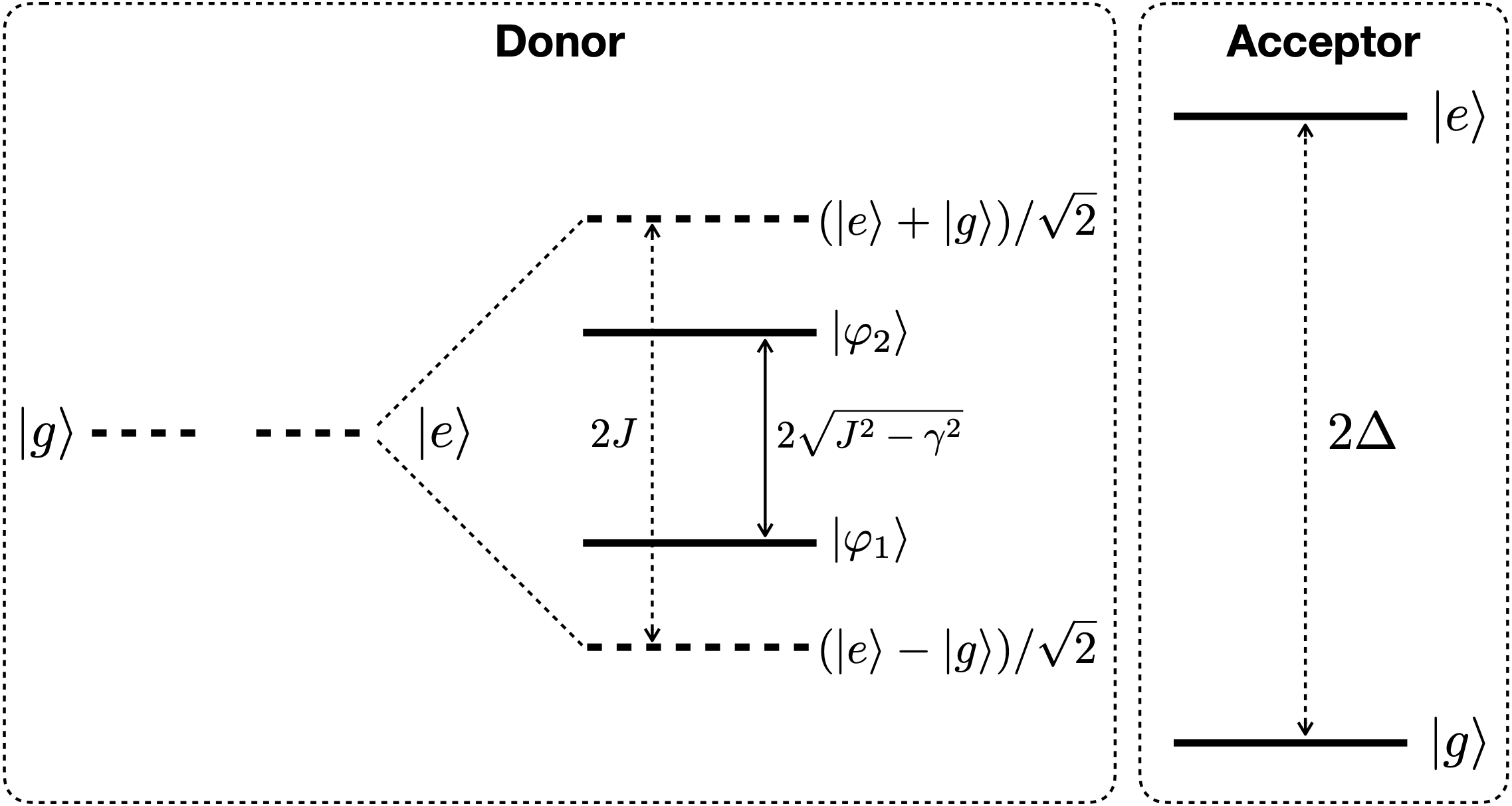}%donoreEnergylevels
\caption{Left panel: The energy levels of the donor in the Hermitian limit (dashed lines) and in the ${\cal PT}$-symmetric regime $J>\gamma$ (solid lines) with the energy gap $2J$ and $2\sqrt{J^2-\gamma^2}$, respectively. The eigenstates $|\varphi_1\rangle=|\tilde{\varphi}_1\rangle/||\tilde{\varphi}_1\rangle|$ and $|\varphi_2\rangle=|\tilde{\varphi}_2\rangle/||\tilde{\varphi}_2\rangle|$ with $\tilde{\varphi}_{1(2)}\rangle$ given in Eq.~(\ref{eq:donorEigenstates}).  Right panel: The energy levels of the acceptor with the energy gap $2\Delta$. Note that the energy levels of decoupled donor and acceptor are also presented in Fig.~\ref{fig:schematic}(a), as grey lines. The parameter values used to generate this figure are $\Delta/J=8$ and $\alpha=\kappa=0$.}
\label{fig:donoreEnergylevels}
\end{figure}

Figure~\ref{fig:donoreEnergylevels} shows the relative energetics of the uncoupled donor and acceptor chromophores. Here dashed lines represent the energy levels of the donor in the Hermitian limit, while solid lines depict the energy levels of the donor in the ${\cal PT}$-symmetric regime. This figure illustrates the primary requirement for uphill transfer, which is $\Delta-J>\alpha/2$, arising from the fact that the energy barrier of $2(\Delta-J)$ between the states $(|e\rangle + |g\rangle)/\sqrt{2}\otimes|g\rangle$ and $(|e\rangle - |g\rangle)/\sqrt{2}\otimes|e\rangle$ should surpass the coherent coupling strength $\alpha$ that links them.

We illustrate the effect of a finite $\Delta$ on the energy spectrum by analyzing the eigenenergies for the decoupled dimer, i.e., $\alpha=0$, resulting in eigenenergies 
%obtained from Eqs.~(\ref{eq:eval_12}) and (\ref{eq:eval_34}) as 
$\lambda_{1,\alpha=0} = -\lambda_{2,\alpha=0} 
= - |\sqrt{J^2-\gamma^2}-\Delta|$ and  
$\lambda_{3,\alpha=0} = - \lambda_{4,\alpha=0} = - |\sqrt{J^2-\gamma^2}+\Delta|$. %\\ 
These eigenenergies are just a sum of monomer energies: $\pm\sqrt{J^2-\gamma^2}$ (donor) and $\pm\Delta$ (acceptor).

\section{Non-Hermitian dimer energy levels and EP characterization for other values of $\Delta/J$}\label{app:dimer}

For completeness, we present here the energetics and EP analysis for two other values of the parameter $\Delta/J$, to compare with the case of $\Delta/J=8$ that favors an uphill transfer for demonstrating the VAET phenomenon discussed in the main text.

\begin{figure}%[hbt!]
\centering
  \includegraphics[width=.99\columnwidth]{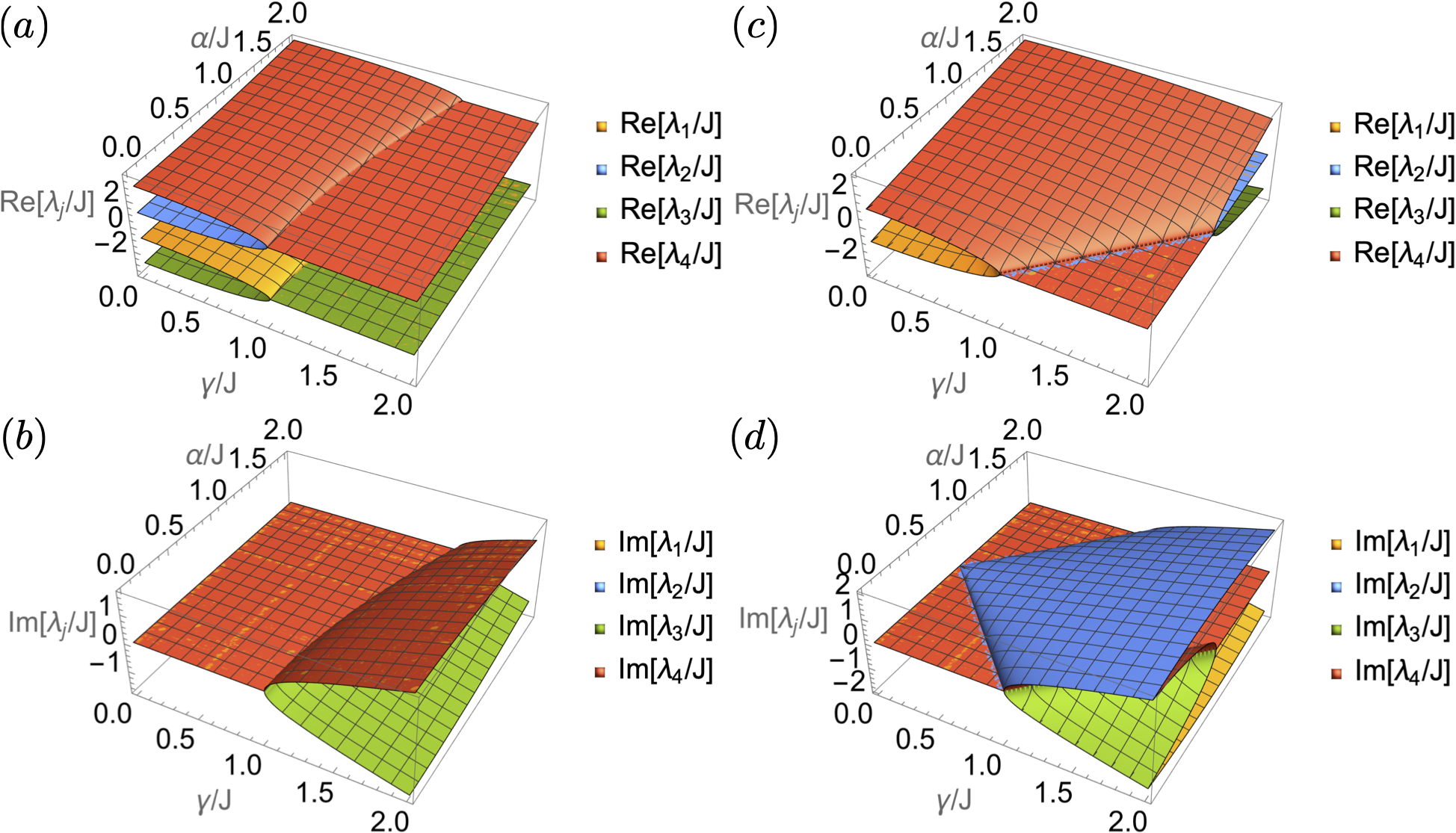}
\caption{(color online) Real and imaginary components of the eigenenergies $\lambda_j$ for $\Delta=2$ in panels (a, b), and for $\Delta/J=0$ in panels (c, d).}
\label{fig:eigenvalue_dimer_z0_z2}
\end{figure}

Figure~\ref{fig:eigenvalue_dimer_z0_z2} presents the real and imaginary components of the eigenenergies $\lambda_j$ as a function of $\gamma$ and $\alpha$, for $\Delta/J=2$ [panels (a, b)] and for $\Delta/J=0$ [panels(c, d)]. 
For weakly uphill $\Delta/J=2$ case in Fig.~\ref{fig:eigenvalue_dimer_z0_z2}(a, b), the energy spectrum exhibits a reduced separation between $\lambda_1, \lambda_3$ and $\lambda_2, \lambda_4$, similar to the $\Delta/J=8$ case in Fig.~\ref{fig:dimer_nonHermitian}(a, b). The non-Hermitian energetics for both of these values host second-order exceptional points. 

In contrast, for the case of $\Delta/J=0$, Fig.~\ref{fig:eigenvalue_dimer_z0_z2}(c, d) shows instead the presence of higher-order exceptional points. 
Physically, when $\Delta=0$, the two levels of the acceptor become degenerate, resulting in a negative energy gap for a downhill excitation transfer from the donor to the acceptor. This implies that the energy transfer in the chromophore dimer does not necessarily require assistance from vibrations in this case.  
The eigenenergies for this case are obtained from Eqs.~(\ref{eq:eval_12}) and (\ref{eq:eval_34}) as
$\lambda_{1,\Delta\rightarrow0} = -\lambda_{2,\Delta\rightarrow0}
=% -\sqrt{\alpha^2-2\alpha J+J^2-\gamma^2} \notag\\
%&=& 
-\sqrt{(\alpha-J)^2 - \gamma^2} 
% \lambda_2(\Delta\rightarrow0) &=& -\lambda_1(\Delta\rightarrow0) = \sqrt{\alpha^2-2\alpha J+J^2-\gamma^2} 
% = \sqrt{(\alpha-J)^2 - \gamma^2}, \\
$ and $\lambda_{3,\Delta\rightarrow0} = - \lambda_{4,\Delta\rightarrow0}
=% -\sqrt{(\alpha^2+2\alpha J+J^2-\gamma^2} \notag\\
%&=& 
-\sqrt{(\alpha+J)^2-\gamma^2} $, %\\
% \lambda_4(\Delta\rightarrow0) &=& -\lambda_3(\Delta\rightarrow0) = \sqrt{\alpha^2+2\alpha J+J^2-\gamma^2} = \sqrt{(\alpha+J)^2-\gamma^2}. 
from which it is clear that for the value of $\Delta=0$, we have a fourth-order EP at $\gamma/J=1$ when $\alpha=0$, as well as second- and third-order EPs as shown explicitly in Figs.~\ref{fig:eigenvalue_dimer_z0_z2}(a) and \ref{fig:eigenvalue_dimer_z0_z2}(b). 
A finite value of $\Delta$, which is associated with the energy splitting of the acceptor chromophore, is  thus critical to lift the degeneracy of the non-Hermitian dimer.

\begin{table}%[hbt!]
\caption{Examples of non-hermiitan dimer eigenenergies $\lambda_j$ for several values of $\gamma$ with $\alpha=1$, $\Delta=8$, and $J=1$.}
\label{tab:numericalEigen}
\centering
\resizebox{0.99\columnwidth}{!}{
\begin{tabular}{|c|c|c|}
\hline %inserts horizontal line
\hline
$\gamma$ & $\{\lambda_1,\lambda_2,\lambda_3,\lambda_4\}$  \\
\hline
$0$ & $\{-7.062, 7.062, -9.062, 9.062\}$   \\
%\hline
%$0.6$ & $\{-7.258, 7.258, -8.865, 8.865\}$  \\
\hline
$0.8$ & $\{-7.453, 7.453, -8.67, 8.67\}$  \\
\hline
$0.9$ & $\{-7.611, 7.611, -8.511, 8.511\}$  \\
\hline
$1.0$ & $\{-7.937, 7.937, -8.185, 8.185\}$  \\
\hline
$1.02$ & $\{-8.061 + 0.156 i, 8.061 - 0.156 i, -8.061 - 0.156 i i, 8.061 + 0.156 i \}$  \\
\hline
$1.04$ & $\{-8.061 + 0.255 i, 8.061 - 0.255 i, -8.061 - 0.255 i, 8.061 + 0.255 i \}$  \\
\hline
$1.06$ & $\{-8.061 + 0.326 i, 8.061 - 0.326 i, -8.061 - 0.326 i, 8.061 + 0.326 i \}$  \\
\hline
$1.08$ & $\{-8.061 + 0.385 i, 8.061 - 0.385 i, -8.061 - 0.385 i, 8.061 + 0.385 i \}$  \\
%\hline
%$1.2$ & $\{-8.06089 + 0.64652 i, 8.06089 - 0.64652 i, -8.06089 - 0.64652 i, 8.06089 + 0.64652 i\}$  \\
%\hline
%$1.6$ & $\{-8.05987 + 1.2335 i, 8.05987 - 1.2335 i, -8.05987 - 1.2335 i, 8.05987 + 1.2335 i\}$  \\
\hline
\end{tabular}
}
\end{table}

Finally, we provide examples of eigenenergies $\lambda_j$ of the non-Hermitian dimer to support the understanding of the results presented in the main text. We consider the Hermitian case $\gamma=0$ as well as values of $\gamma$ near the EP, specifically $\gamma=0.8, 0.9, 1$ or $1.02, 1.04, 1.06, 1.08$ for the ${\cal PT}$-symmetry unbroken or broken phases, respectively, with $J=1$ as the unit. Other parameters used are $\alpha=1$ and $\Delta=8$, consistent with Fig.~\ref{fig:dimer_nonHermitian}(a, b). 
The results are presented in Table~\ref{tab:numericalEigen}. We note that the eigenenergy in the unbroken phase regime is always real, while the transition frequency $\lambda_{13}$ or $\lambda_{42}$ in the broken phase becomes purely imaginary, corresponding to the non-equilibrium steady-state reported in the main text.

\end{document}